\documentclass[%
 reprint,
nofootinbib,
 amsmath,amssymb,
 aps,
showkeys
]{revtex4-2}

\usepackage{graphicx}
\usepackage{dcolumn}
\usepackage{bm}
\usepackage[colorlinks=true,citecolor=magenta,urlcolor=cyan]{hyperref}

\begin{document}

\preprint{APS/123-QED}

\title{Quark spin effects in $e^+e^-$ annihilation: a Monte Carlo event generator study}
\author{A.~Kerbizi$^{a}$}\email{albi.kerbizi@ts.infn.it}
\author{L.~L\"onnblad$^{b}$}\email{leif.lonnblad@fysik.lu.se}
\author{A.~Martin$^{a}$}\email{anna.martin@ts.infn.it}
\affiliation{
$^{a}$Dipartimento di Fisica, Universit\`a  degli Studi di Trieste and INFN Sezione di Trieste,\\
 Via Valerio 2, 34127 Trieste, Italy\\ 
$^{b}$ Department of Physics, Box 118, 221 00 LUND, Sweden}%


\date{\today}

\newcommand\red[1]{{\color{red}#1}}

\def\GeV{\rm GeV}
\def\q{q}
\def\qp{\q'}
\def\qbar{\bar{q}}
\def\qbarp{\bar{q}'}
\def\fL{f_{\rm L}}
\def\thetaLT{\theta_{\rm LT}}
\def\P{\textbf{P}}
\def\PT{P_{\rm T}}
\def\PTi{P_{i\rm T}}
\def\PTa{P_{1\rm T}}
\def\PTb{P_{2\rm T}}
\def\PTzero{P_{0 \rm T}}
\def\PTzeroHat{\hat{\textbf{P}}_{0 \rm T}}
\def\phiH{\phi_{12}}
\def\n{\hat{\textbf{n}}}
\def\Tval{\rm T}
\def\Q{\hat{\textbf{Q}}}
\def\X{X}
\def\Xa{\X_1}
\def\Xb{\X_2}
\def\Xi{\X_i}
\def\AU{A_{12}^{\rm U}}
\def\AL{A_{12}^{\rm L}}
\def\AUL{A_{12}^{\rm UL}}
\def\pmin{\textbf{p}_-}
\def\pp{\textbf{p}_+}
\def\k{k}
\def\kp{k'}
\def\kbar{\bar{k}}
\def\kbarp{\bar{k}'}
\def\kVec{\textbf{k}}
\def\kbarVec{\bar{\textbf{k}}}
\def\xq{\hat{\textbf{x}}_q}
\def\yq{\hat{\textbf{y}}_q}
\def\zq{\hat{\textbf{z}}_q}
\def\xqbar{\hat{\textbf{x}}_{\qbar}}
\def\yqbar{\hat{\textbf{y}}_{\qbar}}
\def\zqbar{\hat{\textbf{z}}_{\qbar}}

\def\sigmaq{\boldsymbol{\sigma}^q}
\def\sigmaqbar{\boldsymbol{\sigma}^{\qbar}}
\def\Iq{1^q}
\def\Iqp{1^{q'}}
\def\sigmaXq{\sigma_x^q}
\def\sigmaYq{\sigma_y^q}
\def\sigmaZq{\sigma_z^q}
\def\Iqbar{1^{\qbar}}
\def\Iqbarp{1^{\qpbar}}
\def\sigmaXqbar{\sigma_x^{\qbar}}
\def\sigmaYqbar{\sigma_y^{\qbar}}
\def\sigmaZqbar{\sigma_z^{\qbar}}
\def\Iden{1}
\def\Trqq{\rm{Tr}_{q\qbar}}

\def\kt{\textbf{k}_{\rm T}}
\def\kpt{\textbf{k}'_{\rm T}}
\def\pt{\textbf{p}_{\rm T}}
\def\ptabs{p_{\rm T}}
\def\kptkpt{\textbf{k}'^2_{\rm T}}

\def\SqT{\textbf{S}_{q\rm T}}
\def\SqbarT{\textbf{S}_{\qbar\rm T}}

\def\ktbar{\bar{\textbf{k}}_{\rm T}}
\def\kptbar{\bar{\textbf{k}}'_{\rm T}}
\def\Pt{\textbf{P}_{\rm T}}
\def\kptbarkptbar{\bar{\textbf{k}}'^2_{\rm T}}

\def\V{\textbf{V}}
\def\VT{\textbf{V}_{\rm T}}
\def\VL{V_{\rm L}}
\def\T{\textbf{T}}
\def\Tr{\rm Tr}

\def\Emin{E_-}
\def\Eq{E_q}

\def\aNN{\hat{a}_{\rm NN}}

\def\Pythia{\textsc{Pythia}}
\def\PythiaUsed{\textsc{Pythia 8.3}}
\def\StringSpinner{\texttt{StringSpinner}}

\def\wh{w_h}
\def\wH{w_H}
\def\VM{\rm VM}
\def\PS{\rm PS}
\def\PSM{\rm PSM}
\def\Im{\rm Im}
\def\Re{\rm Re}
\def\GT{G_{\rm T}}
\def\GL{G_{\rm L}}

\def\AOneTwo{A_{12}}
\def\AOneTwoUL{A_{12}^{UL}}
\def\AOneTwoUC{A_{12}^{UC}}
\def\AOneTwoULC{A_{12}^{UL(UC)}}
\def\Azero{A_0}
\def\AzeroU{A_0^U}
\def\AzeroL{A_0^L}
\def\AzeroUL{A_0^{UL}}
\def\AzeroUC{A_0^{UC}}
\def\AzeroULC{A_0^{UL(UC)}}
\def\HOne{H_{1\q}^{\perp\,h_1}}
\def\HTwo{H_{1\qbar}^{\perp\,h_2}}
\def\DOne{D_{1\q}^{h_1}}
\def\DTwo{D_{1\qbar}^{h_2}}
\def\MOne{m_{h_1}}
\def\MTwo{m_{h_2}}

\def\kperp{\boldsymbol{\kappa}_{1\rm T}}
\def\pperp{\boldsymbol{\kappa}_{2\rm T}}
\def\qT{\textbf{q}_{\rm T}}
\def\QT{Q_{\rm T}}

\def\Hq{H_{1\q}^{\perp\,h}}
\def\Hqbar{H_{1\qbar}^{\perp\,h}}
\def\HqhOne{H_{1\q}^{\perp\,h_1}}
\def\HqhTwo{H_{1\qbar}^{\perp\,h_2}}
\def\Dq{D_{1\q}^{h}}
\def\Ap{a_p^{\q^{\uparrow}\rightarrow h+X}}
\def\ApF{a_p^{F}}
\def\ApU{a_p^{U}}
\def\HF{H_{F}}
\def\HU{H_{U}}
\def\DF{D_{F}}
\def\DU{D_{U}}
\def\phiC{\phi_C}
\def\phiS{\phi_{S_q}}
\def\Hangle{H_{1\q}^{\sphericalangle\,hh}}
\def\fc{f_c}
\def\LL#1{\footnote{LL:~#1}}

\begin{abstract}
Quark spin effects in $e^+e^-$ annihilation to pseudoscalar and vector mesons are implemented for the first time in the \Pythia{} Monte Carlo event generator. The spin-dependent fragmentation of the string stretched between the produced quark-antiquark pair with correlated spin states is described by the string+${}^3P_0$ model implemented in the string fragmentation routine of \Pythia{} by using the \StringSpinner{} package.
The simulated events are used to study the model predictions for the Collins asymmetries of mesons produced back-to-back in the $e^+e^-$ center of mass system by using both the thrust axis method and the hadronic plane method. The obtained asymmetries are compared to the available data from the BELLE and BABAR experiments and the underlying Collins analysing power from the string+${}^3P_0$ model is compared with phenomenological extractions.
\end{abstract}

\maketitle


\section{\label{sec:Introduction} Introduction}
The $e^+e^-$ annihilation to hadrons is a fundamental reaction to study the hadronization, the soft QCD process that converts quarks and gluons to hadrons. According to the factorization theorem \cite{Collins:1981uk}, the $e^+e^-$ reaction can be factorized in the elementary hard interaction $e^+e^-\rightarrow \q\qbar$ where a quark-antiquark pair $\q\qbar$ is produced, and the subsequent hadronization of $\q$ and $\qbar$ in the final state hadrons. The hadronization is described by the fragmentation functions (FFs), which encode the dynamics behind the conversion of quarks and gluons in hadrons, and are thought to be universal. A particularly relevant FF is the Collins function $\Hq$ that implements the Collins effect, namely the fragmentation $\q^{\uparrow}\rightarrow h + X$ of a transversely polarized quark $q$ into unpolarized hadrons \cite{Collins:1992kk}. In the annihilation reaction $e^+e^-\rightarrow h_1\,h_2\,X$, the Collins effect is responsible for the correlations between the azimuthal angles of the hadrons $h_1$ and $h_2$ produced back-to-back in the $e^+e^-$ center of mass system (CMS). The strength of such correlations is quantified by the Collins asymmetry in $e^+e^-$, which couples the functions $\Hq$ and $\Hqbar$. The Collins asymmetry in $e^+e^-$ has been measured to be non-vanishing by the BELLE \cite{Belle:2008fdv,Belle:2019nve}, BABAR \cite{BaBar:2013jdt,BaBar:2015mcn} and BESIII \cite{BESIII:2015fyw} experiments.

In addition to being interesting by itself, the Collins FF is also needed to access the transverse polarization of quarks in a transversely polarized nucleon, encoded in the transversity parton distribution function $h_1^q$. This can be done by measuring the semi-inclusive deep inelastic scattering (SIDIS) process $lN\rightarrow l'\,h\,X$, where a high energy lepton $l$ scatters off a target nucleon $N$ and in the final state at least one hadron $h$ is observed besides the scattered lepton $l'$. If the target nucleon is transversely polarized, $h_1^q$ is coupled with $\Hq$ giving rise to the Collins asymmetries in SIDIS. The Collins asymmetry in SIDIS has been measured by the HERMES experiment using a proton target \cite{HERMES:2010mmo,HERMES:2020ifk}, by the COMPASS experiment using a proton target \cite{COMPASS:2014bze,Alexeev:2022wgr} or a deuteron target \cite{COMPASS:2023vhr}, and by the HALL A experiment at the JeffersonLab facility using a neutron target \cite{JeffersonLabHallA:2011ayy}. The asymmetry has been measured to be nonvanishing for a proton target showing that both $h_1^q$ and $\Hq$ are different from zero.

The Collins asymmetries in SIDIS and in $e^+e^-$ have been used by several groups in combined phenomenological analyses aimed at extracting $h_1^q$ and $\Hq$ at the same time \cite{Anselmino:2015sxa,Martin:2014wua,Anselmino:2015fty,Kang_transv_evolution,Cammarota:2020qcw}. As a result of the analyses $h_1^q$ and $\Hq$ are given in terms of chosen parametrizations, the free parameters of which are obtained from fits to the Collins asymmetries in SIDIS and $e^+e^-$. The resulting parametrization for $\Hq$ represents our knowledge on this FF.

An alternative approach to the phenomenological extractions using parametrizations of FFs is the modeling of the spin effects in hadronization and the implementation of the model in Monte Carlo (MC) event generators (MCEGs). Models represent our understanding of the physical mechanisms involved in hadronization, and MCEGs are needed to perform calculations that allow the comparison between the model predictions and the data. 
As explained in the following, work in this direction started recently. This paper, focused on the quark spin effects in $e^+e^-$ annihilation, represents a further step forward in making realistic this alternative approach.

A recently developed model is the string+${}^3P_0$ model, which is an extension of the Lund Model of string fragmentation \cite{Andersson:1983ia} implemented in \Pythia{} \cite{Bierlich:2022pfr}. The model includes the quark spin degree of freedom at the amplitude level and was extensively studied by using standalone MC implementations~\cite{Kerbizi:2018qpp,Kerbizi:2019ubp,Kerbizi:2021M20}. More recently, it was implemented in the hadronization part of the \Pythia{} MCEG (the \StringSpinner{} package \cite{Kerbizi:2023cde}) for the simulation of the polarized deep inelastic scattering with production of pseudo-scalar mesons (PSMs) \cite{Kerbizi:2021StringSpinner} and vector mesons (VMs) \cite{Kerbizi:2023cde}. \StringSpinner{} was used to carry simulations of SIDIS with a transversely polarized proton target and to study the model results for the transverse spin asymmetries like the Collins asymmetry and the dihadron asymmetries. A promising agreement with data was found \cite{Kerbizi:2023cde}.

In this paper we describe the first implementation of the quark spin effects in the \Pythia{} \texttt{8.3} MCEG in the simulation of $e^+e^-$ annihilation to hadrons taking into account the correlations between the spins of the intermediate $\q\qbar$ pair.
To describe the quark spin effects in the string fragmentation we use the development of the string+${}^3P_0$ model for the fragmentation of a string stretched between a $\q\qbar$ pair with correlated spin states~\cite{Kerbizi:2023luv}. The implementation in \Pythia{} is achieved by extending the \StringSpinner{} package of Ref.~\cite{Kerbizi:2023cde} to $e^+e^-$ annihilation events including the spin effects for the production of final state PSMs and VMs. We use the new package to simulate $e^+e^-$ events at the CMS energy $\sqrt{s}=10.6\,\GeV$ corresponding to the energy of the BELLE \cite{Belle:2008fdv} and BABAR \cite{BaBar:2013jdt} experiments.

The detailed description of the implementation of $e^+e^-$ annihilation with quark spin effects in \Pythia{} is presented in Sec.~\ref{sec:Implementation}. In Sec.~\ref{sec:asymmetries} we summarize the formalism of the Collins asymmetries measured in $e^+e^-$. The results on such asymmetries from simulated $e^+e^-$ annihilation events are discussed in Sec.~\ref{sec:Results} and compared with the results from the BELLE and BABAR experiments. In Sec.~\ref{sec:analysing-power} we calculate the Collins analysing power, the ratio of $\Hq$ and the spin-averaged FF $\Dq$, from the string+${}^3P_0$ model and compare it with phenomenological extractions. Finally, the conclusions are given in Sec.~\ref{sec:Conclusions}.

\section{\label{sec:Implementation} Implementation of the string+${}^3P_0$ model in \Pythia{} for $e^+e^-$ annihilation}

In this section we describe in detail the different steps applied in \StringSpinner{} to implementat the quark spin effects in \Pythia{} for $e^+e^-$ annihilation to hadrons. The starting points are the \StringSpinner{} package in Ref.~\cite{Kerbizi:2023cde} and the string+${}^3P_0$ model in Ref.~\cite{Kerbizi:2023luv}.

The simulation of $e^+e^-$ annihilation consists of three main steps: the generation of the kinematics associated to the hard reaction $e^+e^-\rightarrow \q\qbar$, the construction of the joint spin density matrix of the $\q\qbar$ pair, and the hadronization $\q\qbar\rightarrow h_1,h_2,\dots,$ by fragmenting the string stretched between $\q$ and $\qbar$ in the final state hadrons $h_1, h_2, \dots$. Gluon radiation as simulated by the final state parton shower has been switched off, since would produce strings with gluon ``kinks'' between the quark and anti-quark with correlated spin states, and such configurations are not yet handled by the string+${}^3P_0$ model. Further final state effects such as the Bose-Einstein correlations are also not included in the string+${}^3P_0$ model, and have been switched off.

\subsection{The hard reaction $e^+e^-\rightarrow \q\qbar$}
\begin{figure}[th]
\centering
\begin{minipage}[b]{0.45\textwidth}
\hspace{-2.0em}
\includegraphics[width=1.0\textwidth]{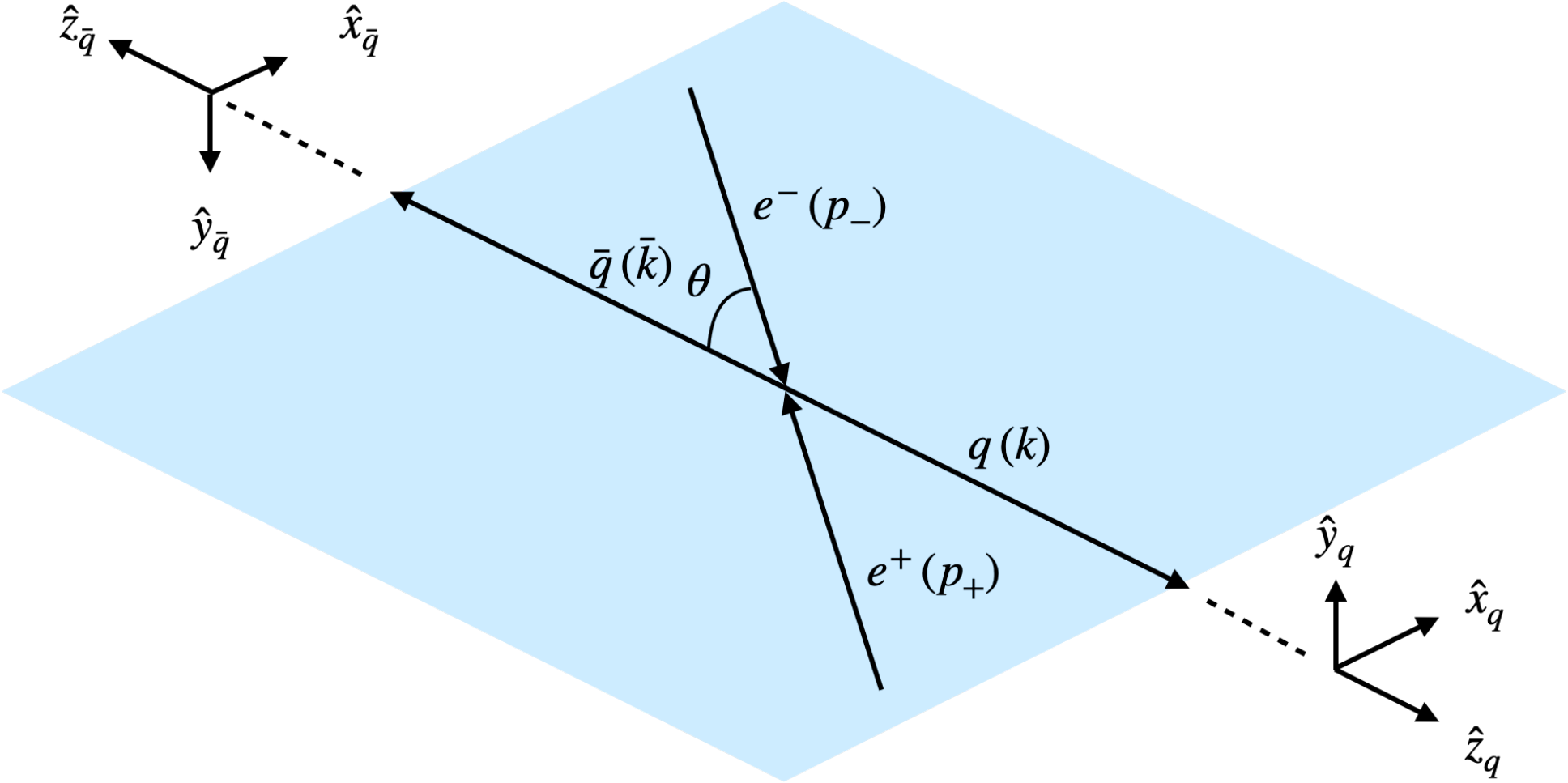}
\end{minipage}
\caption{Kinematics of the $e^+e^-\rightarrow q\qbar$ in the CMS.}
\label{fig:kinematics}
\end{figure}

To begin the simulation, we let \Pythia{} generate the hard reaction $e^+e^-\rightarrow \q\qbar$, assuming the annihilation is mediated by a virtual photon $\gamma^*$. The flavor of the quark $q$ is selected among the kinematically allowed flavours in proportion to the squared charges $e^2_q/\sum_a e^2_a$. For the generation of the kinematics, the differential cross section associated to $e^+e^-\rightarrow \q\qbar$ is used. The kinematics in the $e^+e^-$ CMS is shown in Fig.~\ref{fig:kinematics}a, where $\theta$ is angle between the momentum $\pmin$ of $e^-$ and the momentum $\kVec$ of $\q$. The momenta of $e^+$ and $\qbar$ are indicated by $\pp=-\pmin$ and $\kbarVec=-\kVec$, respectively. 

Following Ref.~\cite{Kerbizi:2023luv} 
, we define the quark helicity frame (QHF) by the set of axes $\lbrace \xq, \yq,\zq\rbrace$. The axes are obtained as $\zq=\kVec/|\kVec|$, $\yq=\pmin\times\zq/|\pmin\times \zq |$ and $\xq= \yq\times \zq$. The antiquark helicity frame (AHF) is defined by the axes $\lbrace \xqbar, \yqbar, \zqbar\rbrace$, obtained analogously by using $\kbarVec$ instead of $\kVec$. The QHF and AHF are also shown in Fig.~\ref{fig:kinematics}. They coincide with the definitions used in Ref.~\cite{DAlesio:2021dcx}. In the $e^+e^-$ CMS, the momenta of $e^-$ and $q$ can be expressed in the QHF as $\pmin=\sqrt{s}\,(-\sin\theta,0,\cos\theta)/2$ and $\kVec=\sqrt{s}\,(0,0,1)/2$, where the electron and the quark masses are neglected. 

\subsection{The joint spin density matrix of $\q\qbar$}
Once the $\q\qbar$ pair is generated, the joint spin density matrix $\rho(\q,\qbar)$ is set up. It implements the correlations between the (entangled) spin states of $\q$ and $\qbar$. Neglecting quark masses, the joint spin density matrix reads \cite{Chen:1994ar,Kerbizi:2023luv}
\begin{equation}\label{eq:rho}
\rho(\q,\qbar)=\frac{1}{4}\left[\Iq\otimes \Iqbar - \sigmaZq\otimes \sigmaZqbar + \aNN\,(\sigmaXq\otimes \sigmaXqbar + \sigmaYq\otimes \sigmaYqbar)\right],
\end{equation}
where $\sigma_{i}^{\q(\qbar)}$ indicates the Pauli matrix along the axis $i=x,y,z$ in the QHF (AHF), and $1^{q(\qbar)}$ is the identity matrix. The quantity $\aNN(\theta)=\sin^2\theta/(1+\cos^2\theta)$ describes the correlation between the transverse spin states of $q$ and $\qbar$ originated by the tensor polarization of the $\gamma^*$.

The spin density matrices of $q$ and $\qbar$ are obtained from the joint spin density matrix as
\begin{eqnarray}\label{eq:rho q}
    \rho(q)=\Tr_{\qbar}\,\rho(\q,\qbar), \,\,\,\,\,\,\,\,\,\,\,\, \rho(\qbar)=\Tr_{\q}\,\rho(\q,\qbar).
\end{eqnarray}
Inserting Eq.~(\ref{eq:rho}) in Eq.~(\ref{eq:rho q}), it is $\rho(q)=\Iq/2$ and $\rho(\qbar)=\Iqbar/2$, meaning that $q$ and $\qbar$ are not separately polarized. Rather, their spin states are correlated. 
%
%
\begin{figure}[tbh]
\centering
\begin{minipage}[b]{0.45\textwidth}
\hspace{1.0em}
\includegraphics[width=1.0\textwidth]{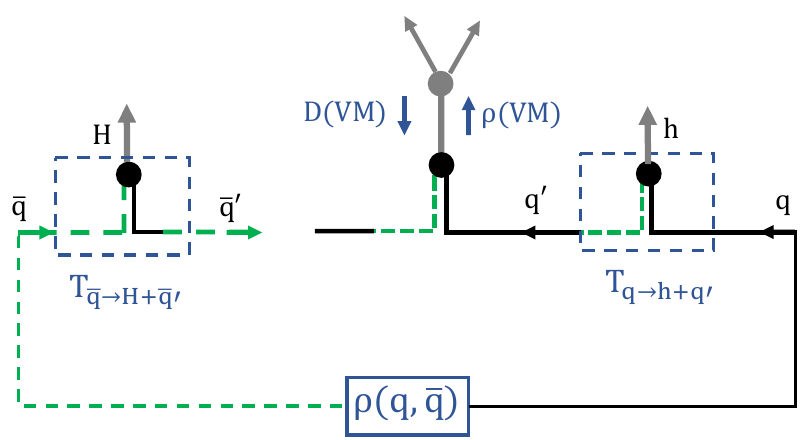}
\end{minipage}
\caption{Representation of the polarized string fragmentation process in \StringSpinner{}.}
\label{fig:string fragmentation}
\end{figure}

\subsection{The string fragmentation of the $\q\qbar$ pair}
The string fragmentation of the $\q\qbar$ pair is simulated by \Pythia{} as a recursive process of elementary quark splittings $q\rightarrow h +\qp$ and elementary antiquark splittings $\qbar\rightarrow H+\qbarp$, as shown in Fig.~\ref{fig:string fragmentation}. The splittings are taken from the $\q$ or the $\qbar$ side randomly with equal probability.

In the quark splitting $q\rightarrow h+\qp$ the emitted hadron $h$ has four-momentum $p$, while the leftover quark $\qp$ has four-momentum $\kp$. Momentum conservation yields $p=\k-\kp$, where $\k$ is the four-momentum of the fragmenting quark $q$. The transverse momenta of $\q$, $h$ and $\qp$ with respect to the string axis (i.e. the $\q\qbar$ relative momentum in the CMS) are indicated by $\kt$, $\pt$ and $\kpt$, respectively. They are related by $\pt=\kt-\kpt$.

In the antiquark splitting $\qbar\rightarrow H + \qbarp$, the emitted hadron $H$ has four-momentum $P$ while the leftover antiquark $\qbarp$ has four-momentum $\kbarp$. Four-momentum conservation implies $P=\kbar-\kbarp$, where $\kbar$ is the four-momentum of the fragmenting antiquark $\qbar$. The transverse momenta of $\qbar$, $H$ and $\qbarp$ with respect to the string axis are defined as $\ktbar$, $\Pt$ and $\kptbar$, respectively. They are related by $\kptbar = \ktbar-\Pt$. 

The $h$ and $H$ mesons are restricted to be PSMs and VMs, since only these are present in the string+${}^3P_0$ model of Ref.~\cite{Kerbizi:2023luv}.

To implement the spin effects for an $e^+e^-$ annihilation event, we start from the previous implementation of StringSpinner \cite{Kerbizi:2023cde}
and use the string+${}^3P_0$ model for $e^+e^-$ annihilation in Ref.~\cite{Kerbizi:2023luv}. The description of the involved steps in the simulation of $e^+e^-$ annihilation in \Pythia{} is as follows.

\subsubsection{Splitting from the $\q$ side}\label{sec:splitting q side} Let us suppose the first splitting is taken from the $\q$ side. In the string+${}^3P_0$ model the splitting $\q\rightarrow h+\qp$ is described by the $2\times 2$ splitting matrix \cite{Kerbizi:2023luv}
\begin{equation}\label{eq:Tq}
    T_{\qp,h,q} = (\dots) \left[\mu+\sigma_z^q\boldsymbol{\sigma}^{q}\cdot \kpt\right]\times\Gamma(h).
\end{equation}
The dots indicate the scalar term of the splitting amplitude describing the energy-momentum sharing between $h$ and $\qp$, already implemented in \Pythia{}. The quantity $\mu=\Re(\mu)+i\,\Im(\mu)$ is the complex free parameter called “complex mass", accounting for the ${}^3P_0$ state of quark-antiquark pairs produced at the string breakups. 
The vector $\boldsymbol{\sigma}^q=(\sigmaXq, \sigmaYq, \sigmaZq)$ is the vector of Pauli matrices in the QHF.
The matrix $\Gamma(h)$ describes the coupling of $q$ and $\qp$ with $h$. It is $\Gamma(h)=\sigmaZq$ for $h=\PSM$, and $\Gamma(h)=\GT\,\sigmaq\sigmaZq\cdot\VT^{*}+\Iq\,\GL\,\VL^*$ for $h=\VM$. The vector $\V=(\VT,\VL)$ is the linear polarization of the VM in the QHF. The free parameters $\GT$ and $\GL$ describe the coupling of $\q$ and $\qp$ with a transversely and a longitudinally polarized VM, respectively.

To introduce the spin effects in the splitting $q\rightarrow h+\qp$ according to the string+${}^3P_0$ model, the hadron $h$ emitted by \Pythia{} is rejected if it is not a PSM or a VM. A new one is thus generated by \Pythia{}, which is accepted with the probability
\begin{eqnarray}\label{eq:wh} \wh(\kpt;\SqT)&=&\frac{\Trqq\left[\T^a_{\qp,h\,\q}\,\rho(\q,\qbar)\T^{a\,\dagger}_{\qp,h\,\q}\right]}{\Trqq\left[\T^b_{\qp,h\,\q}\,\T^{b\,\dagger}_{\qp,h\,\q}\right]} \\
\nonumber &=& \frac{1}{2}\,\left[1+c\,\frac{2\,\Im(\mu)}{|\mu|^2+\kptkpt}\,\SqT\cdot\left(\zq\times\kpt\right)\right],
\end{eqnarray}
where $\T^{a}_{\qp,h,\q}=T^{a}_{\qp,h,\q}\otimes \Iqbar$. Here the splitting amplitude for VM emission is written as $\T_{\qp,h,\q}=\T^a_{\qp,h,\q}\,\V^*_a$, with $a=\xq, \yq, \zq$ labelling the linear polarization state of the VM in the QHF. A summation over the repeated indices is understood. If $h=\PSM$, the indices $a,b$ are omitted.
$\wh$ can be interpreted as the ratio between the probabilities for a polarized and an unpolarized splitting $\q\rightarrow h+\qp$ in the string+${}^3P_0$ model.

The second line in Eq.~(\ref{eq:wh}) is obtained by using the expression for the splitting matrix in Eq.~(\ref{eq:Tq}). The effect of $\wh$ is to introduce correlations between the transverse momentum $\pt$ of $h$ and the transverse polarization $\SqT$ of $\q$, namely the transverse part of $\textbf{S}_q=\Tr\,\sigmaq\,\rho(q)$ [see Eq.~(\ref{eq:rho q})]. It thus changes the azimuthal distribution of $h$ produced by \Pythia{} to emulate the spin effects of the string+${}^3P_0$ model and, for a non-zero $\SqT$, it is responsible for the Collins effect in the emission of $h$. The factor $c$ is $-1$ for a PSM and $\fL=|\GL|^2/(2|\GT|^2+|\GL|^2)$ for a VM, and governs the relative sign of the Collins effect for PSM and VM emissions. The parameter $\fL$ describes the fraction of longitudinally polarized VMs.

Equation (\ref{eq:wh}) is the analogue of the probability introduced in Refs.~\cite{Kerbizi:2021StringSpinner, Kerbizi:2023cde} for the description of the spin effects in the DIS process.

\subsubsection{Decay of vector mesons}\label{sec:decay}
If $h$ is a PSM, it does not carry spin information and the decay is handled by \Pythia{}. If $h$ is a VM, its decay is instead handled by \StringSpinner. In this case the spin density matrix of the VM is used for the simulation of the decay process. The (not normalized) spin density matrix reads \cite{Kerbizi:2023luv}
\begin{eqnarray}\label{eq:rho VM}
    \rho_{aa'}(h)\propto \Trqq\left[\T^{a}_{\qp,h\,\q}\,\rho(\q,\qbar)\T^{a'\,\dagger}_{\qp,h\,\q}\right].
\end{eqnarray}

The angular distribution of the decay products in the rest frame of the VM is generated according to $dN(p\rightarrow p_1, p_2,..)/d\Phi(p_1,p_2,..)\propto \rho_{aa'}\,\hat{M}_a(p\rightarrow p_1, p_2,..)\,\hat{M}_{a'}(p\rightarrow p_1, p_2,..)$, where $\hat{M}_{a}(p\rightarrow p_1,p_2,..)$ is the amplitude describing the decay of a VM with linear polarization $a$ in the daughters $d_1,d_2,..$, and $d\Phi(p_1,p_2,..)$ indicates the relevant differential phase space factor.

The decay of $h$ returns the decay matrix $D_{a'a}=\hat{M}_{a'}\,\hat{M}_a$ \cite{Kerbizi:2023luv}. This matrix is required by the Collins-Knowles recipe \cite{Collins:1987cp,Knowles:1988vs} to propagate the information on the orientation of the decay hadrons to the leftover quark $\qp$, as schematically shown in Fig.~\ref{fig:string fragmentation}.

\subsubsection{Propagation of the spin correlations}\label{sec:propagation} The spin correlations are propagated by calculating the joint spin density matrix $\rho(\qp,\qbar)$ of the remaining $\qp\qbar$ pair. The (not-normalized) spin density matrix is given by \cite{Kerbizi:2023luv}
\begin{equation}\label{eq:rho q'qb}
    \rho(\qp,\qbar)\propto \T^{a}_{\qp,h,\q}\,\rho(\q,\qbar)\T^{a'\dagger}_{\qp,h,\q}\,D_{a'a}.
\end{equation}
For a VM emission the decay matrix $D_{a'a}$ is required, while for a PSM emission the indices $a$ and $a'$, and $D_{a'a}$ are removed.
The matrix $\rho(\qp,\qbar)$ now contains the information on the emission of $h$ from the quark side.

\subsubsection{Splitting from the $\qbar$ side}\label{sec:splitting qbar side}
After the first splitting, the next one is taken by \Pythia{} randomly either from the $\qp$ side or the $\qbar$ side, with equal probability. Let us suppose it is 
$\qbar\rightarrow H+\qbarp$ from the $\qbar$ side. The procedure for implementing the spin effects is analogous to that of the $\q$ side. 

The spin-dependent antiquark splitting is described in the string+${}^3P_0$ model by the splitting amplitude $\T_{\qbarp,H,\qbar}=\Iq\otimes T_{\qbar',H,\qbar}$, where \cite{Kerbizi:2023luv}
\begin{equation}\label{eq:Tqb}
    T_{\qbarp,H,\qbar} = (\dots)\,\left[\mu+\sigmaZqbar\boldsymbol{\sigma}^{\qbar}\cdot \kptbar\right]\times\Gamma(H).
\end{equation}
The meaning of the different terms composing $T_{\qbarp,H,\qbar}$ is analogous to those in Eq.~(\ref{eq:Tq}), with the difference that the Pauli matrices $\sigmaqbar=(\sigmaXqbar,\sigmaYqbar,\sigmaZqbar)$, the transverse momentum $\kptbar$ as well as the polarization vector entering $\Gamma(H)$ for $H=\VM$ are expressed in the AHF.

To introduce the spin effects in the splitting $\qbar\rightarrow H+\qbarp$, the hadron $H$ generated by \Pythia{} is rejected if it is not a PSM or a VM, and a new one is generated by \Pythia{} still from the $\qbar$ side of the string. The latter is accepted with probability
\begin{eqnarray}\label{eq:wH}
w_H(\kptbar;\SqbarT)&=&\frac{\Tr_{\qp\qbar}\left[\T^a_{\qbarp,H\,\qbar}\,\rho(\qp,\qbar)\T^{a\,\dagger}_{\qbarp,H\,\qbar}\right]}{\Tr_{\rm \qp\qbar}\left[\T^{b}_{\qbar',H\,\qbar}\,\T^{b\,\dagger}_{\qbarp,H\,\qbar}\right]} \\
\nonumber &=& \frac{1}{2}\,\left[1+c\,\frac{2\,\Im(\mu)}{|\mu|^2+\kptbarkptbar}\,\SqbarT\cdot\left(\zqbar\times\kptbar\right)\right],
\end{eqnarray}
where the second line is obtained by using Eq.~(\ref{eq:Tqb}).

$\wH$ introduces a modulation in the azimuthal distribution of the transverse momentum $\Pt$ of $H$ expressed in the AHF. The modulation depends in this case on the transverse part $\SqbarT$ of the polarization vector $\textbf{S}_{\qbar}=\Tr\,\sigmaqbar\,\rho(\qp,\qbar)$ of $\qbar$.
As shown in Ref.~\cite{Kerbizi:2023luv}, after the emission of $h$ the antiquark $\qbar$ acquires a transverse polarization $\SqbarT\neq \textbf{0}$ depending on the transverse momentum $\pt$\footnote{The same is true if first a hadron $H$ is emitted from $\qbar$, then $h$ is emitted from $q$. In this case the quark $q$ acquires a transverse polarization $\SqT\neq \textbf{0}$ that depends on the transverse momentum $\Pt$ of $H$.}. Through Eq.~(\ref{eq:wH}), this leads to correlations between the azimuthal angles of the transverse momenta $\pt$ of $h$ and $\Pt$ of $H$.
$\wH$ can therefore be interpreted as the conditional probability for emitting $H$ from the $\qbar$ side of the string once the hadron $h$ has been emitted from the $\q$ side. In the string+${}^3P_0$ model, this is the mechanism for the generation of the Collins asymmetry for hadrons produced back-to-back in $e^+e^-$ \cite{Kerbizi:2023luv}.

For the decay of $H$ and the propagation of the spin information after its generation, the same steps as in Sec.~\ref{sec:decay} and Sec.~\ref{sec:propagation} are employed provided that the substitutions $\T_{\qp,h,\q}\rightarrow \T_{\qbarp,H,\qbar}$ and $\rho(\q,\qbar)\rightarrow \rho(\qp,\qbar)$ are performed. 


\subsubsection{Exit condition.}
The procedure described in the previous paragraphs is applied recursively by randomly emitting hadrons from the $q$ and $\qbar$ sides of the string, until the exit condition of the string fragmentation process is called by \Pythia{}. At this step a remaining string piece $\q_m\,\qbar_n$ must be fragmented by one last breaking by a $\q'\,\qbar'$ pair and therefore the production of the final two hadrons $h=\q_m\qbar'$ and $H=\q'\qbar_n$. 

To handle this last step, if the previous splitting was taken from the
antiquark side, we treat the production of $h$ as the splitting
$\q_m\rightarrow h+\q'$ followed by the projection of the $\q'\qbar_n$
state onto the hadronic state $H$. This leads to a reweight procedure
for $h$ as in Sec.~\ref{sec:splitting q side} and to the decay
according to Sec.~\ref{sec:decay} ($H$ is taken to be unpolarized). If
$h$ is not accepted by this procedure, \Pythia{} rejects the full
fragmentation chain and starts all over again \footnote{This is a
feature of the standard \Pythia{}. By changing the code manually in such a way that only the final two hadrons are rejected, we checked
that the simulation results do not differ to a noticeable degree. A
gain in the execution time of the simulations is however observed.}.

Analogously, if the previous splitting was taken from the quark side, the production of $H$ is seen as the splitting $\qbar\rightarrow H+\qbarp$ and treated following Sec.~\ref{sec:splitting qbar side}. In this case, if $h$ is a VM, it is taken to be unpolarized.

This recipe is somewhat simplified with respect to that proposed in Ref.~\cite{Kerbizi:2023luv}. We checked, however, that the simulation results obtained with the two recipes do not differ to a noticeable degree.
This is due to the fact that the spin information decays along the fragmentation chain, and the possible spin effects in the production of the final two hadrons are negligible.

\section{The Collins asymmetries in $e^+e^-$}\label{sec:asymmetries}
In inclusive two-hadron production in $e^+e^-$ annihilation, $e^+e^-\rightarrow h_1\,h_2\,X$, two Collins asymmetries are introduced. They are based on two different methods: the thrust axis method, which leads to the asymmetry $\AOneTwo$, and the hadronic plane method, which leads to the asymmetry $\Azero$. The methods exploit different reference planes for the measurement of the relevant azimuthal angles and lead to different theoretical expressions for the Collins asymmetries.

A quantity used in both methods is the thrust axis, the best approximation of the $\q\qbar$ axis that is measurable. It is defined as the normalized vector $\n$ that maximises the event variable $T=\sum_j\,|\P_j\cdot\n|/\sum_j\,|\P_j|$. The index $j$ runs over the final state hadrons, and $\P_j$ is the momentum of the hadron $h_j$ in the CMS. $T$ is referred to as the thrust, and it is $T\leq 1$. $T\simeq 1$ indicates a two-jet like configuration, while $T\simeq 0.5$ indicates that the distribution of the produced hadrons is roughly spherical. In experimental analyses, the two-jet like events are selected by requiring $\Tval>0.8$ \cite{Belle:2008fdv,BaBar:2013jdt}.

The thrust axis $\n$ is used to select the back-to-back hadrons $h_1$ and $h_2$ forming a pair produced in the same event, by requiring $(\P_1\cdot \n)\,(\P_2\cdot \n)<0$. To reduce the association of hadrons to the wrong hemisphere, the selection $\QT<3.5\,\GeV/c$ is applied. $\QT$ is the magnitude of the photon transverse momentum evaluated in the rest frame of the $h_1\,h_2$ pair \cite{Boer:1997mf,Belle:2008fdv}.
 The requirements on the thrust and the photon transverse momentum are a common feature of the analyses performed by the BELLE \cite{Belle:2008fdv,Belle:2019nve} and BABAR \cite{BaBar:2013jdt,BaBar:2015mcn} collaborations. We apply them in this paper as well.


\begin{figure}[tbh]
\centering
\begin{minipage}[b]{0.45\textwidth}
\hspace{1.0em}
\includegraphics[width=1.0\textwidth]{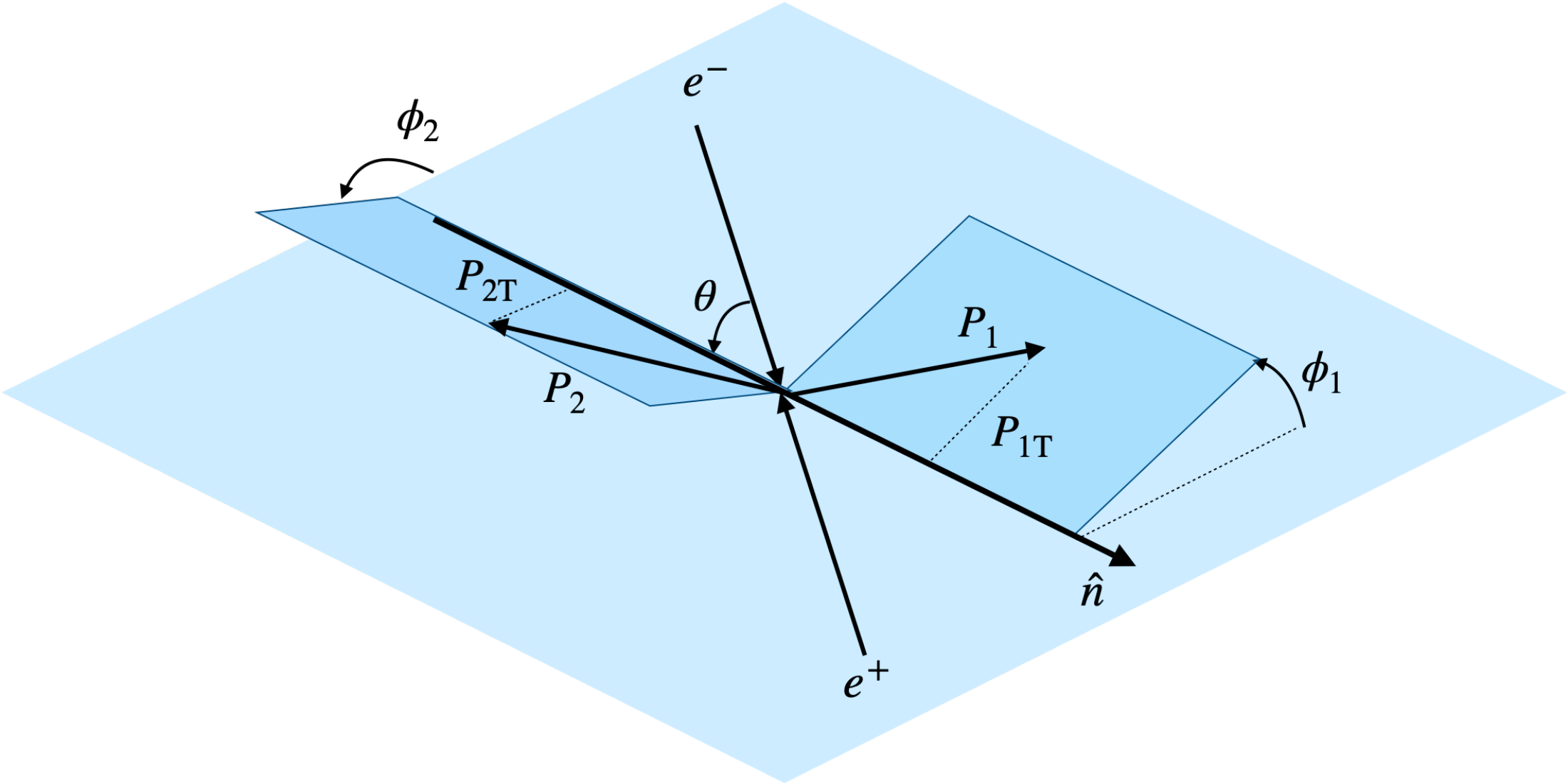}
\end{minipage}
\caption{The kinematics associated to the thrust axis method.}
\label{fig:kinematic M12}
\end{figure}

\subsubsection{The $\AOneTwo$ asymmetry}\label{sec: A12 theory}
For the evaluation of the $\AOneTwo$ asymmetry, a reference plane containing the momentum $\pmin$ of the beam $e^-$ and the thrust axis $\n$ is used to measure the azimuthal angles in the CMS. As shown in Fig.~\ref{fig:kinematic M12},
the transverse momentum of $h_i$ with respect to $\n$ is indicated by $\P_{i\rm T}$ and the corresponding azimuthal angle is indicated by $\phi_i$, where $i=1,2$.

The distribution of the azimuthal angle $\phiH = \phi_1+\phi_2$ is then considered. It is expected to be (see, e.g., Ref.~\cite{Boer:2008fr,Anselmino:2015sxa,DAlesio:2021dcx})
\begin{eqnarray}\label{eq:dN}
\nonumber    N_{12}&&(\phiH;z_1,z_2,\PTa,\PTb)\propto \\ &&1+\frac{\langle\sin^2\theta\rangle}{\langle 1+\cos^2\theta\rangle}\,A_{12}(z_1,z_2,\PTa,\PTb)\,\cos\phiH,
\end{eqnarray}
where the fractional energy of $h_i$ is defined as $z_i=2\,E_i/\sqrt{s}$, with $E_i$ the energy of $h_i$.
The amplitude $\AOneTwo$ depends on the fractional energies $z_1$, $z_2$ and on the transverse momenta $\PTa$ and $\PTb$. 
The dependence of $\AOneTwo$ on the hard scale $Q^2=s$ is omitted in Eq.~(\ref{eq:dN}).
The partonic expression of $\AOneTwo$ reads
\begin{eqnarray}\label{eq:A12}
\AOneTwo = \frac{\sum_{q,\qbar}\,e_q^2\,\frac{\PTa}{z_1\MOne}\HOne\,\frac{\PTb}{z_2\MTwo}\HTwo}{\sum_{q\qbar}\,e_q^2\,\DOne\, \DTwo},
\end{eqnarray}
where $e_q$ is the charge of $q$ in units of the elementary charge and $m_{h_i}$ is the mass of hadron $h_i$. The numerator in Eq.~(\ref{eq:A12}) involves the product of two Collins FFs $\HOne(z_1,\PTa)$ and $\HTwo(z_2,\PTb)$ describing the fragmentations of transversely polarized $q$ and $\qbar$, respectively. The summation over the flavors and antiflavors expresses the fact that each hadron can be either produced in the fragmentation of $q$ or of $\qbar$. 
The denominator in Eq.~(\ref{eq:A12}) is given by the product of the spin-averaged FFs $\DOne(z_1,\PTa)$ and $\DTwo(z_2,\PTb)$, which describe the fragmentations of the unpolarized $\q$ and $\qbar$. 

The asymmetry $\AOneTwo$ is not the directly measured quantity. To eliminate systematic effects originated by false asymmetries, the angular distribution in Eq.~(\ref{eq:dN}) is used to construct the normalized yields $R_{12}(\phiH)=N_{12}(\phiH)/\langle N_{12}\rangle$ in a chosen kinematic bin, where $\langle N_{12}\rangle$ is the average yield in that bin. The ratios $R_{12}^U$, $R_{12}^L$ and $R_{12}^C$ are evaluated using unlike-sign (U) and like-sign (L) pairs, while (C) use any pair of charged hadrons.
Finally the ratios
\begin{eqnarray}\label{eq:A12UL}
\nonumber R_{12}^{UL(UC)}=\frac{R_{12}^U}{R_{12}^{L(C)}}\simeq 1+\frac{\langle \sin^2\theta\rangle}{\langle 1+\cos^2\theta\rangle}\,\AOneTwoULC\,\cos\phiH,\,\,\,\,\,\,\,\,\, \\
\end{eqnarray}
are used to measure the amplitudes $\AOneTwoULC$. They are given by
\begin{equation}
\AOneTwoULC\simeq A_{12}^{U}-A_{12}^{L(C)},
\end{equation}
that is by the difference between the Collins asymmetry in Eq.~(\ref{eq:A12}) for unlike sign hadrons ($\AOneTwo^U$) and the asymmetry for like sign ($\AOneTwo^L$) or charged ($\AOneTwo^C$) hadrons.

Note that in Eq.~(\ref{eq:A12UL}) we use the convention that the factor $\langle\sin^2\theta\rangle/\langle1+\cos^2\theta\rangle$ is not included in the definition of $\AOneTwoULC$, at variance with the definition employed experimentally \cite{Belle:2019nve,Belle:2008fdv,BaBar:2013jdt,BaBar:2015mcn}. 

\subsubsection{The $\Azero$ asymmetry}\label{sec:A0 theory}
\begin{figure}[tbh]
\centering
\begin{minipage}[b]{0.45\textwidth}
\hspace{1.0em}
\includegraphics[width=1.0\textwidth]{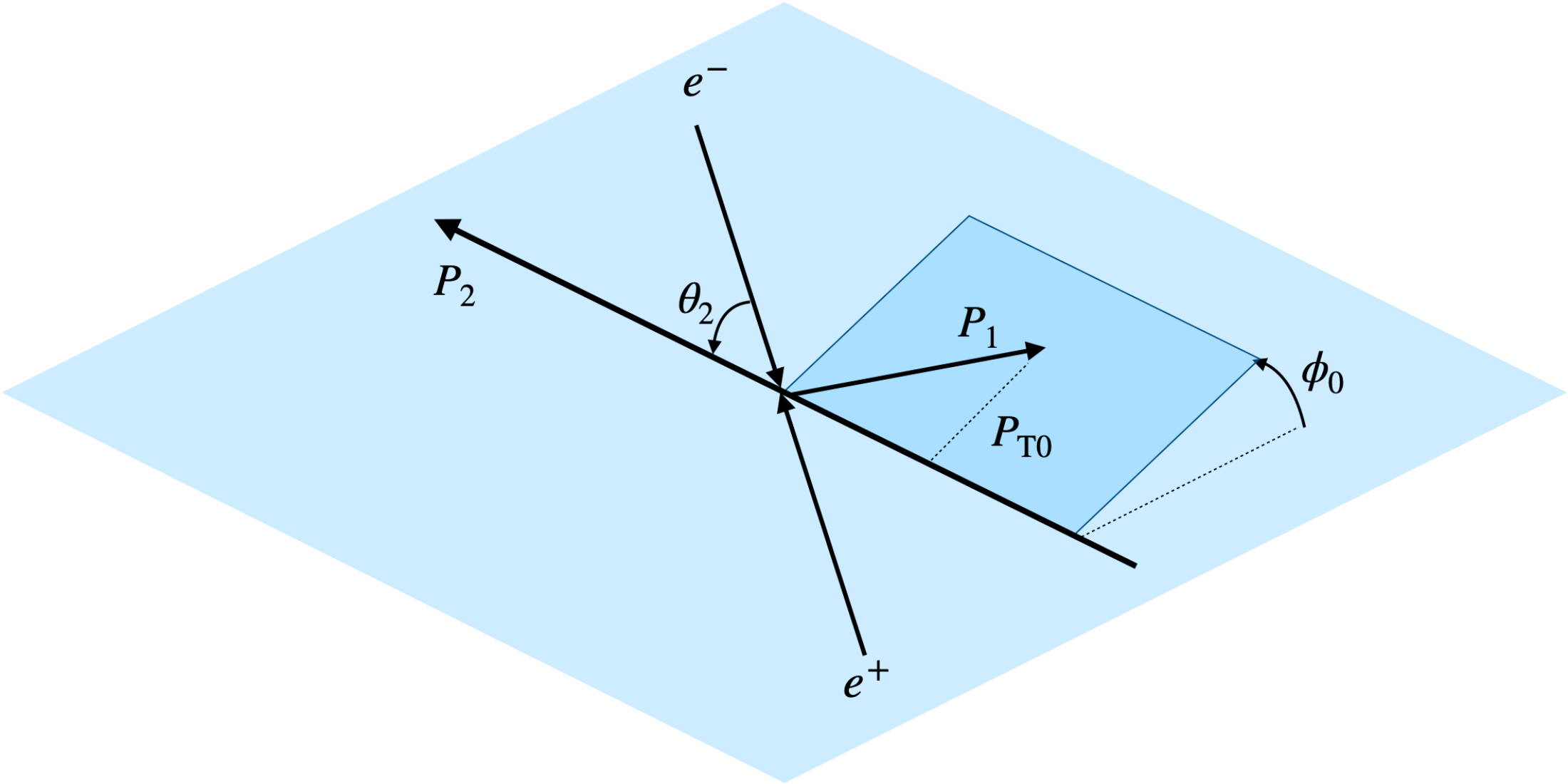}
\end{minipage}
\caption{The kinematics associated to the hadronic plane method.}
\label{fig:kinematic M0}
\end{figure}

To evaluate the asymmetry $\Azero$, the plane containing the momentum $\pmin$ of $e^-$ and the momentum $\P_2$ of $h_2$ is considered, as shown in Fig.~\ref{fig:kinematic M0}. The plane is used to measure the azimuthal angle $\phi_0$ of the transverse momentum $\PTzero$ of $h_1$ with respect to $\P_2$.

The distribution of the azimuthal angle $\phi_0$ of $h_1$ is expected to be~\cite{Boer:1997mf,Boer:2008fr,Anselmino:2015sxa,DAlesio:2021dcx}
\begin{eqnarray}\label{eq:N0}
\nonumber     N_{0}&&(\phi_0;z_1,z_2,\PTzero)\propto \\ &&1+\frac{\langle\sin^2\theta_2\rangle}{\langle 1+\cos^2\theta_2\rangle}\,A_{0}(z_1,z_2,\PTzero)\,\cos 2\phi_0,
\end{eqnarray}
where $\theta_2$ is the angle between $\P_2$ and the beam $\pmin$. The amplitude $A_0$ depends on $z_1$, $z_2$, $\PTzero$ and $\QT$~\cite{Boer:1997mf}. The latter dependence is not explicitly shown. The partonic expression of $A_0$ is \cite{Boer:1997mf,Boer:2008fr}
\begin{equation}\label{eq:A0}
  \Azero = \frac{\sum_{q,\qbar}\,e_q^2\,\mathcal{C}\left[w\,\HOne\,\HTwo\right]}{\sum_{q,\qbar}\,e_q^2\,\mathcal{C}\left[\DOne\, \DTwo\right]}.
\end{equation}
At variance with the $\AOneTwo$ asymmetry in Eq.~(\ref{eq:A12}), $\Azero$ is given in terms of the convolution of FFs over the involved transverse momenta. The convolution integral $\mathcal{C}$ is defined as
\begin{eqnarray}
\nonumber \mathcal{C}\left[w\,H_1\,H_2\right]=&&\int\,d^2\kperp\,d^2\pperp\,\delta^{(2)}(\kperp+\pperp-\qT)\times\\
&&w(\kperp,\pperp)\,H_1(z_1,\kperp^2)\,H_2(z_2,\pperp^2),
\end{eqnarray}
for two generic FFs $H_{1,2}$ and weight factor $w$. The weight $w$ appearing in the numerator in Eq.~(\ref{eq:A0}) can be written as $w(\kperp,\pperp)=2\,\textbf{P}_{0\rm T}\cdot\kperp\,\textbf{P}_{0\rm T}\cdot\pperp/\PTzero^2-\kperp\cdot\pperp$. The integral is performed over the transverse momenta $\kperp$ of $\q$ and $\pperp$ of $\qbar$ in the rest frame of the pair $h_1\,h_2$. Transverse momentum conservation in the photon decay to the quark pair is ensured by the $\delta$-function, with $\qT=-\textbf{P}_{0\rm T}/z_1$ being the photon transverse momentum in the rest frame of the $h_1h_2$ pair. 

Similarly to the $\AOneTwo$ asymmetry, $\Azero$ is measured from the ratio $R_0^{UL(UC)}$ between the normalized yields $R_0^U=N_0^U/\langle N_0^U\rangle$ for unlike sign hadrons and $R_0^{L(C)}=N_0^{L(C)}/\langle N_0^{L(C)}\rangle$ for like sign (L) or charged (C) hadrons. The ratios
\begin{eqnarray}\label{eq:A0UL}
\nonumber R_{0}^{UL(UC)}=\frac{R_{0}^U}{R_{0}^{L(C)}}\simeq 1+\frac{\langle\sin^2\theta_2\rangle}{\langle 1+\cos^2\theta_2\rangle}\,\AzeroULC\,\cos 2\phi_0, \\
\end{eqnarray}
are used to measure $\AzeroUL$, the Collins asymmetry with the hadronic plane method. In terms of the asymmetry in Eq.~(\ref{eq:A0}), it is
\begin{equation}
    \AzeroULC\simeq \AzeroU-\Azero^{L(C)}.
\end{equation}
Likewise to Eq.~(\ref{eq:A12UL}), $\AzeroULC$ is given by the difference between the Collins asymmetries for unlike charged ($\Azero^U$) and like charged ($\Azero^L$) or charged ($\Azero^C$) back-to-back hadron pairs.

Note that, as can be seen from Eq.~(\ref{eq:A0UL}), we use the convention that the factor $\langle\sin^2\theta_2\rangle/\langle 1+\cos^2\theta_2\rangle$ is not included in the definition of the $\AzeroULC$ asymmetry.

\section{The Collins asymmetries from simulated $e^+e^-$ events}\label{sec:Results}
To study the quark-spin effects in $e^+e^-$ annihilation in the string+${}^3P_0$ model we have evaluated both the $\AOneTwo$ and $\Azero$ Collins asymmetries following the data analysis described in the previous section.

The settings and values of the relevant parameters used in the simulations are described in Sec.~\ref{sec:setting}.

In Sec.~\ref{sec:Thrust Method} we present the results on the asymmetry $\AOneTwo$ and compare them with the results of the BELLE and BABAR experiments.

The corresponding results for $\Azero$ are given in Sec.~\ref{sec:Hadronic Method}.

\subsection{Simulation settings and validation}\label{sec:setting}
We generated $60\cdot10^{\,6}$ $e^+e^-$ annihilation events with \Pythia{} and the new \StringSpinner{} package. The events have been generated at the same energy for BELLE and BABAR, namely $\sqrt{s}=10.6\,\GeV$. The annihilation reaction is mediated by a virtual photon, which is allowed to decay to $\q\qbar$ pairs with $q=u,d,s$. Thus the production of the heavier charm and bottom quarks has been switched off.

For the free parameters of the (spin-less) Lund string Model we used the default values in \PythiaUsed{}, as we did in Ref.~\cite{Kerbizi:2023cde}.  The additional free parameters are those introduced by the string+${}^3P_0$ model. For the complex mass $\mu$ we use $\Re(\mu)=0.42\,\GeV/c^2$ and $\Im(\mu)=0.76\,\GeV/c^2$ as in Ref.~\cite{Kerbizi:2023cde}. The fraction $\fL$ of longitudinally polarized VMs with respect to the string axis is set to $\fL=0.12$, meaning that VM production with transverse polarization with respect to the string axis is favored. The oblique polarization $\thetaLT$ is taken to be $\thetaLT=-0.65$, meaning that the interference between VMs with longitudinal and transverse polarizations with respect to the string axis is different from zero. The values of $\fL$ and $\thetaLT$ differ from those used in Ref.~\cite{Kerbizi:2023cde}, $\fL=0.92$ and $\thetaLT=0$. The values used here were selected to obtain a satisfactory agreement with the experimental results. They also give a satisfactory comparison with the Collins and dihadron asymmetries in SIDIS, which were the observables considered in Ref.~\cite{Kerbizi:2023cde}.

\subsection{Results on the $\AOneTwo$ asymmetry}\label{sec:Thrust Method}
As explained in Sec.~\ref{sec:asymmetries}, the $\AOneTwo$ asymmetry can be measured using the thrust axis as an approximation of the $\q\qbar$ axis. This introduces a smearing in the measured angles of the final state hadrons and in the measured asymmetries. The correction can be evaluated using a MCEG as done for the BELLE results published in 2008 \cite{Belle:2008fdv} and the BABAR results published in 2014 and 2015 \cite{BaBar:2013jdt,BaBar:2015mcn}. The other option is not to correct for the use of the thrust axis, as in the case of the BELLE results published in 2019 \cite{Belle:2019nve}. Both sets of data are discussed in the following.

\subsubsection{$\AOneTwo$ asymmetry with thrust axis correction}\label{sec:Results qq axis}
Being the experimental asymmetries corrected for the use of the thrust axis, the asymmetries from the simulated events have been calculated using the true $\q\qbar$ axis. In the analysis of simulated events, the thrust axis is used only to form the pair using hadrons from opposite hemispheres and to apply the selection on $\Tval$. For each simulated event we use the \Pythia{} routine for the event analysis to find the thrust axis $\n$ and to evaluate the corresponding value of the thrust $\Tval$.

For each charged hadron in a given hepisphere, all possible pairs are formed using all the back-to-back charged hadrons. In each kinematical bin, the distribution of the azimuthal angle $\phiH$ for the unlike sign pairs, like sign pairs and charged pairs is constructed.
Finally, the ratios $R_{12}^{UL(UC)}$ are fitted with a fit function based on Eq.~(\ref{eq:A12UL}) and the Collins asymmetries $\AOneTwoUL$ and $\AOneTwoUC$ are extracted.

\begin{figure}[tbh]
\centering
\begin{minipage}[b]{0.5\textwidth}
\includegraphics[width=1.0\textwidth]{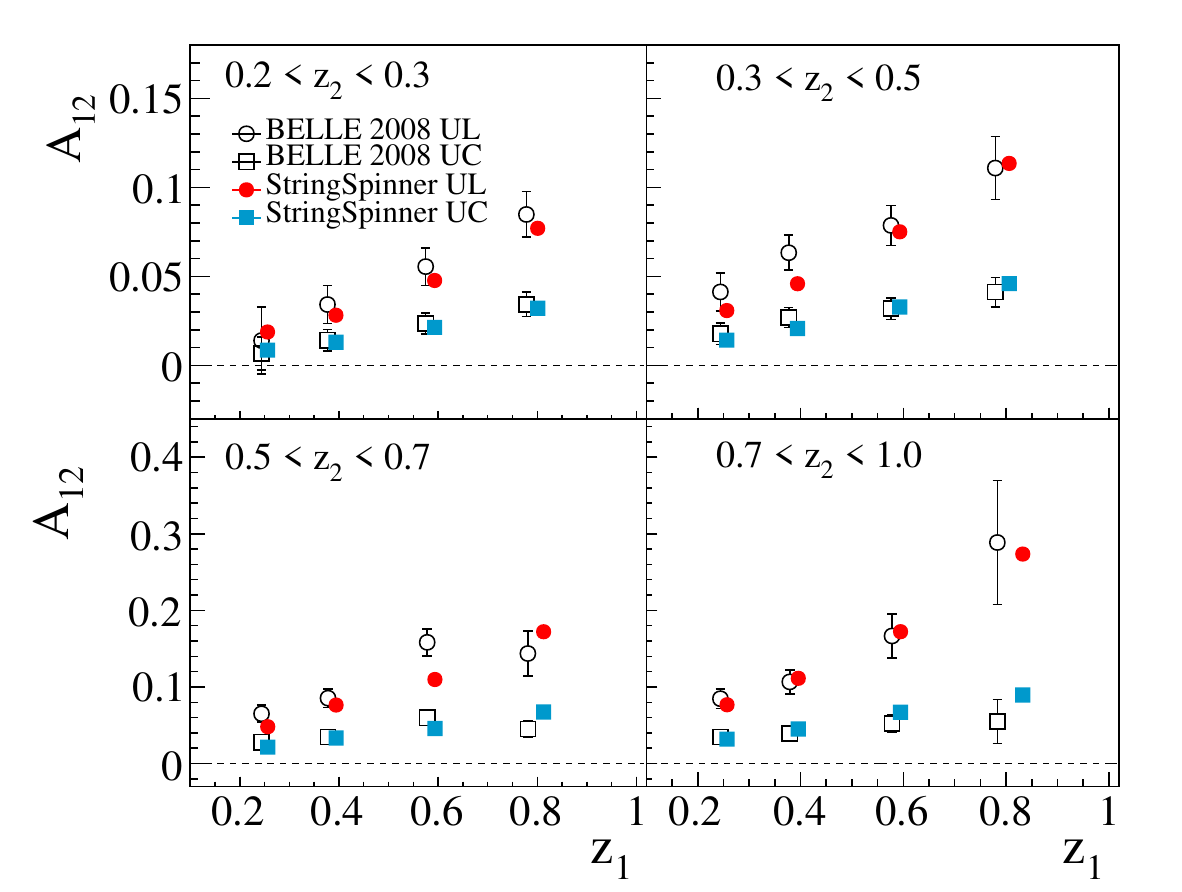}
\end{minipage}
\caption{Comparison between the Collins asymmetries $\AOneTwoUL$ and $\AOneTwoUC$ for back-to-back $\pi^{\pm}-\pi^{\mp}$ pairs obtained with \StringSpinner{} (full markers) and the corresponding asymmetries measured by BELLE \cite{Belle:2008fdv} (open markers).}
\label{fig:A12 Belle 2008}
\end{figure}

\paragraph{Comparison with $\AOneTwo$ results from BELLE.}
In Fig.~\ref{fig:A12 Belle 2008} we show the results for the asymmetries $\AOneTwoUL$ (full circles) and $\AOneTwoUC$ (full squares) for charged pions pairs obtained with \StringSpinner{}. The asymmetries are given as a function of $z_1$ for different bins of $z_2$. The binning is the same as in the BELLE analysis \cite{Belle:2008fdv}. Each hadron is required to have $z_i>0.2$. 

\begin{figure}[thb]
\centering
\begin{minipage}[b]{0.5\textwidth}
\includegraphics[width=1.0\textwidth]{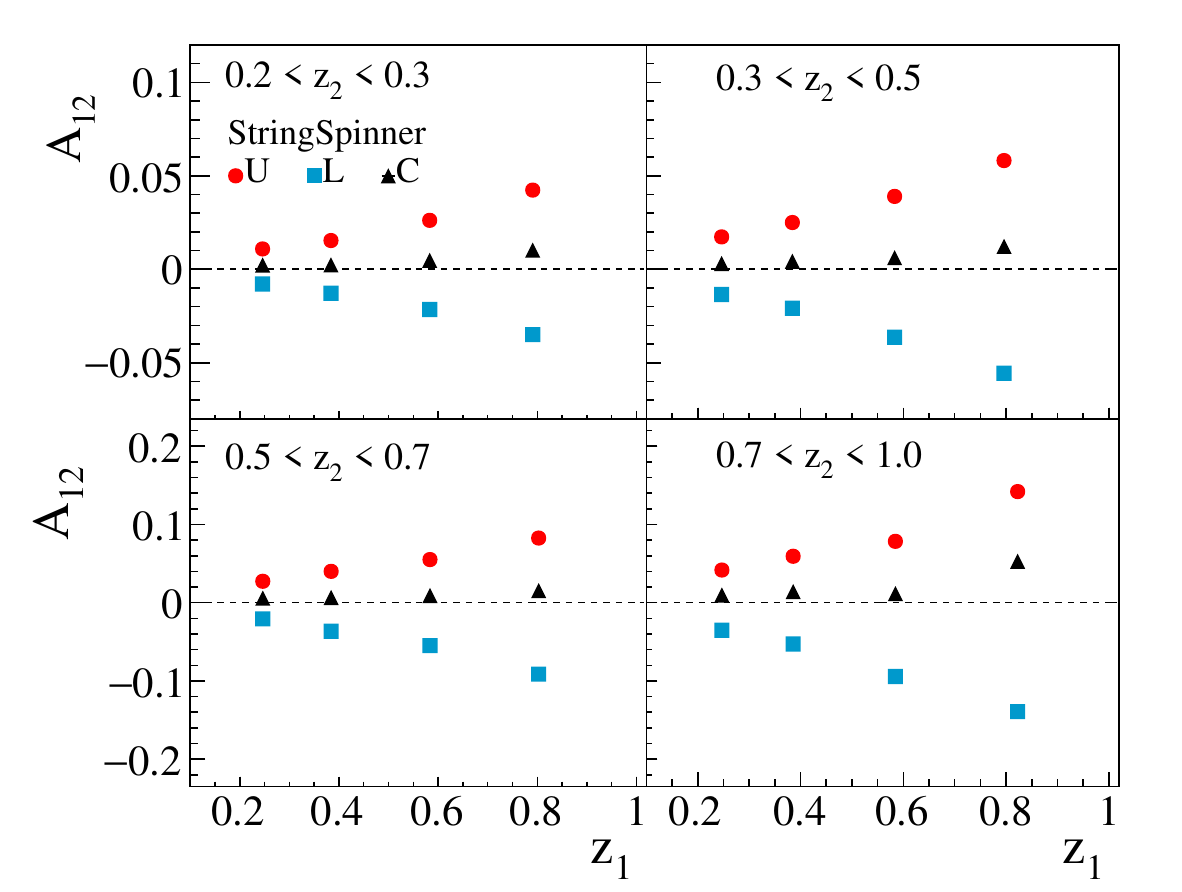}
\end{minipage}
\caption{The Collins asymmetries $\AOneTwo^U$ (circles), $\AOneTwo^L$ (squares) and $\AOneTwo^C$ (triangles) for back-to-back $\pi^{\pm}-\pi^{\mp}$ pairs as obtained with \StringSpinner{}. The same simulated events as for the asymmetries in Fig.~\ref{fig:A12 Belle 2008} have been used.}
\label{fig:A12ULC }
\end{figure}

As can be seen, the simulated asymmetries show a rising trend as a function of $z_1$ in each $z_2$ bin. Hadrons with large fractional energies are likely to be produced close to the initial quarks, where the correlations between the spin states of $\q$ and $\qbar$ are strongest and the resulting Collins effect is large. As studied in detail in Refs.~\cite{Kerbizi:2018qpp,Kerbizi:2019ubp,Kerbizi:2021M20}, the spin information in the fragmentation chain decays as long as more hadrons are produced (with small $z$ values), resulting in a small Collins effect. Note also that the simulations reproduce smaller $\AOneTwoUC$ asymmetries than the $\AOneTwoUL$ asymmetries, as observed in the data.

To better understand how the $\AOneTwoUL$ and $\AOneTwoUC$ asymmetries arise in the string+${}^3P_0$ model, in Fig.~\ref{fig:A12ULC } we show separately the asymmetries $\AOneTwo^U$ (circles), $\AOneTwo^L$ (rectangles) and $\AOneTwo^C$ (triangles). As can be seen, $\AOneTwo^U$ is positive while $\AOneTwo^L$ is negative. This is expected since the product of two favored Collins FFs contribute to $\AOneTwo^U$, while the product of a favored and an unfavored Collins FF appears in $\AOneTwo^L$, and in the string+${}^3P_0$ model the favoured and unfavoured Collins FFs have opposite signs. $\AOneTwo^C$ is instead close to zero due to the fact that both U and L pairs contribute, which have Collins asymmetries with opposite signs.

In Fig.~\ref{fig:A12 Belle 2008} we also show the asymmetries $\AOneTwoUL$ (open circles) and $\AOneTwoUC$ (open squares) measured by BELLE \cite{Belle:2008fdv}. The errorbars associated to the data correspond to the total uncertainty obtained by summing in quadrature the statistical and the systematic uncertainties. The same is done also for the data considered in the following. Note that the BELLE data are already corrected for the contribution of charm quarks \cite{Belle:2008fdv}. Thus the comparison with the MC results is consistent. As can be seen, the comparison with data is satisfactory. The model reproduces the size and the trends of both the experimental $\AOneTwoUL$ and $\AOneTwoUC$ asymmetries. 

Note that the simulated asymmetries have been slightly shifted horizontally by a constant amount to better show the comparison with data. This is done also in the following.

\begin{figure}[tbh]
\centering
\begin{minipage}[b]{0.5\textwidth}
\includegraphics[width=1.0\textwidth]{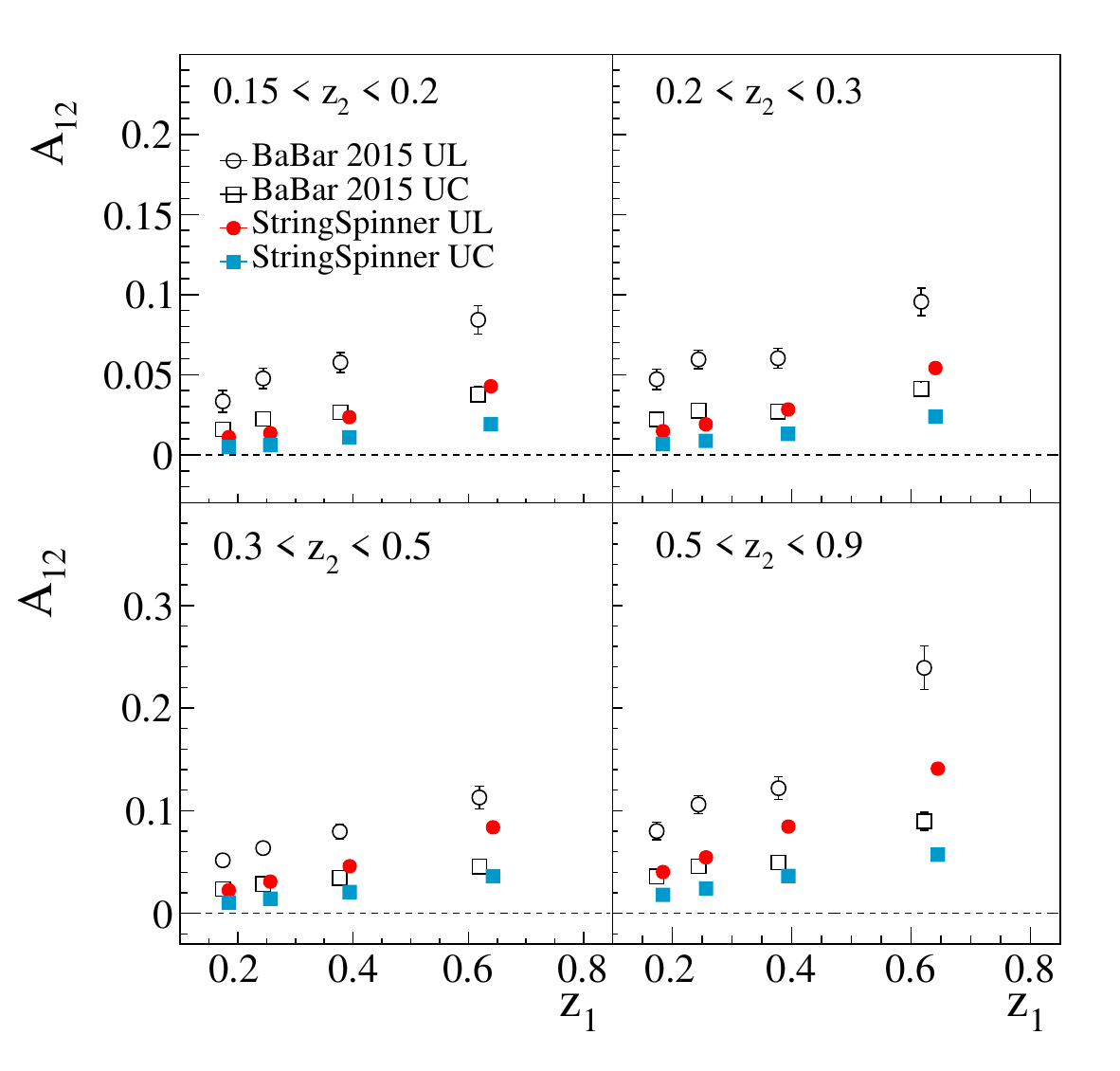}
\end{minipage}
\caption{The Collins asymmetries $\AOneTwoUL$ (full circles) and $\AOneTwoUC$ (full squares) for back-to-back $\pi^{\pm}-\pi^{\mp}$ pairs as obtained with \StringSpinner{}, and as measured by BABAR \cite{BaBar:2015mcn} (open markers).}
\label{fig:A12 BaBar 2014}
\end{figure}

\paragraph{Comparison with $\AOneTwo$ from BABAR.}
The BABAR collaboration has also measured the $\AOneTwo$ asymmetries for back-to-back charged pion pairs as a function of the fractional energy $z_i$ with a slightly different binning \cite{BaBar:2015mcn}. The asymmetries $\AOneTwoUL$ (open circles) and $\AOneTwoUC$ (open squares) measured by BABAR are shown in Fig.~\ref{fig:A12 BaBar 2014}. They are corrected for the smearing effect caused by the misalignment between $\n$ and the $\q\qbar$ axis as well as for the contribution of charm quarks. The $z_i$ range has been set to $0.15<z_i<0.9$. In addition a cut $\alpha_i<\pi/4$ has been applied, where the opening angle $\alpha_i$ is the opening angle of $h_i$ with respect to the thrust axis~\cite{BaBar:2015mcn}.

The closed points in Fig.~\ref{fig:A12 BaBar 2014} show the results obtained from the MC simulations as a function of the fractional energy. The trend is the same as in Fig.~\ref{fig:A12 Belle 2008}. In particular, they show a rising trend as a function of the fractional energy. The size of the asymmetries turns out to be smaller as compared to the experimental results. This is somehow expected since results from BELLE and BABAR are different, as already noted in Ref.~\cite{BaBar:2013jdt}.

\begin{figure}[tbh]
\centering
\begin{minipage}[b]{0.5\textwidth}
\includegraphics[width=1.0\textwidth]{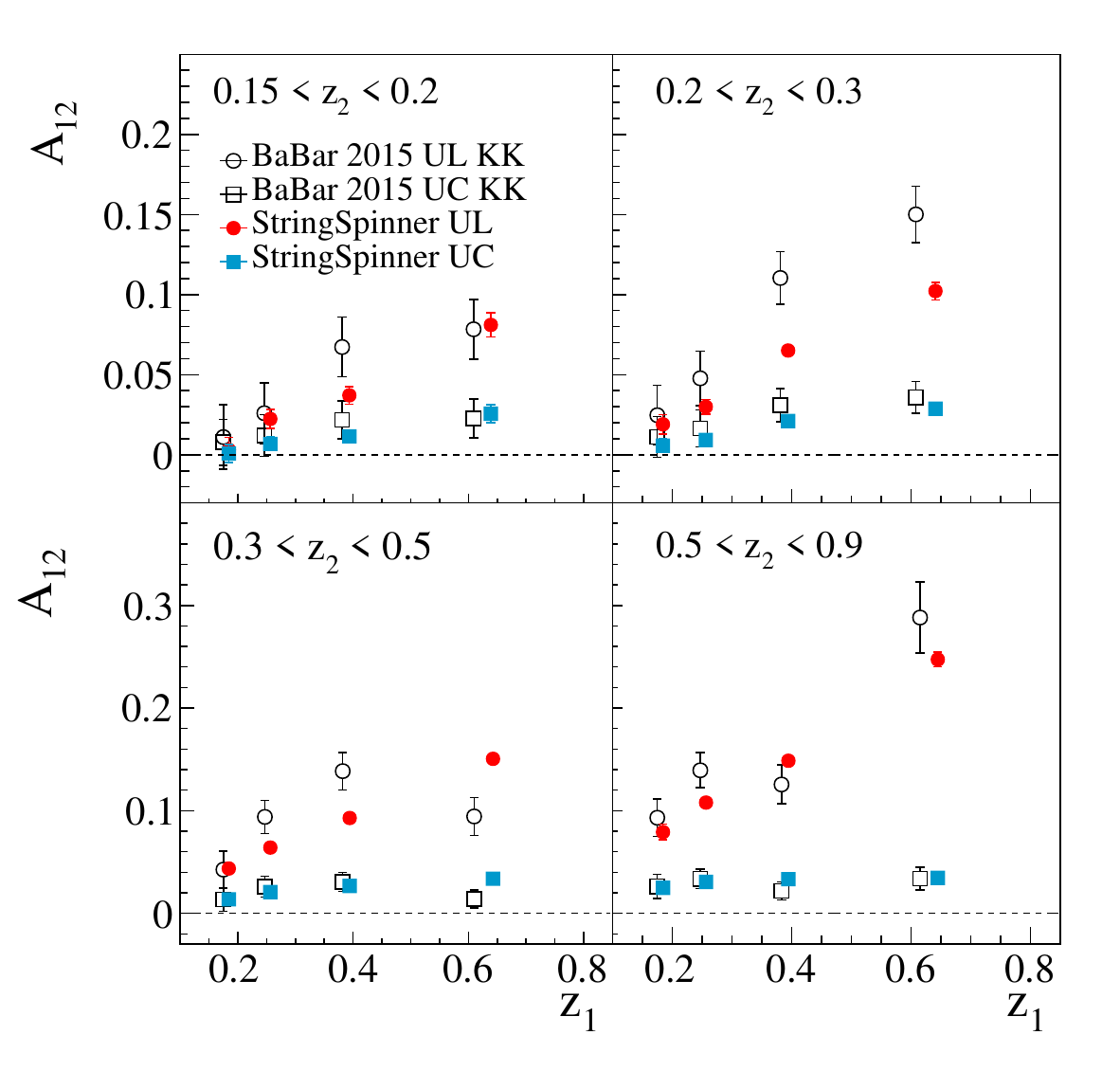}
\end{minipage}
\caption{The Collins asymmetries $\AOneTwoUL$ (full circles) and $\AOneTwoUC$ (full squares) for back-to-back $K^{\pm}-K^{\mp}$ pairs as obtained with \StringSpinner{}, and as measured by BABAR \cite{BaBar:2015mcn} (open markers).}
\label{fig:A12 KK BaBar 2015}
\end{figure}

In Fig.~\ref{fig:A12 KK BaBar 2015} we show the Collins asymmetries $\AOneTwoUL$ (full circles) and $\AOneTwoUC$ (full rectangles) for charged kaon pairs from MC events. The asymmetries are shown as function of $z_1$ for the selected bins of $z_2$, and each kaon is required to have a fractional energy $0.15<z_i<0.9$ and an opening angle $\alpha_i<\pi/4$ as in the BABAR analysis \cite{BaBar:2015mcn}. 
Comparing with the $\AOneTwo$ asymmetries for charged pions in Fig.~\ref{fig:A12 BaBar 2014}, one can see that the string+${}^3P_0$ model produces a larger Collins asymmetry for kaons. To understand this feature, we recall that in the string+${}^3P_0$ model with only PSM production, the Collins asymmetry for the production of pions is similar to the asymmetry for the production of kaons~\cite{Kerbizi:2018qpp}. When introducing also VM production and decay, the decay products of VMs contribute to a dilution of the Collins asymmetry of the final hadrons~\cite{Kerbizi:2021M20}. The dilution is less for kaons as compared to pions, due to the fact that the fraction of mesons from decays of VMs is lower in the kaon sample than in the pion sample.

The corresponding Collins asymmetries for charged kaons measured by BABAR are shown in Fig.~\ref{fig:A12 KK BaBar 2015} by the open markers. As can be seen, \StringSpinner{} satisfactorily reproduces the size of the measured asymmetries for small and large $z$. Note that for kaon pairs the reduction of the $\AOneTwoUC$ asymmetries with respect to the $\AOneTwoUL$ asymmetries is more pronounced than in the pion case (cf. with Fig.~\ref{fig:A12 Belle 2008}), which is reproduced by the simulations.

\begin{figure}[!h]
\centering
\begin{minipage}[b]{0.42\textwidth}
\hspace{-0.8em}
\includegraphics[width=1.0\textwidth]
{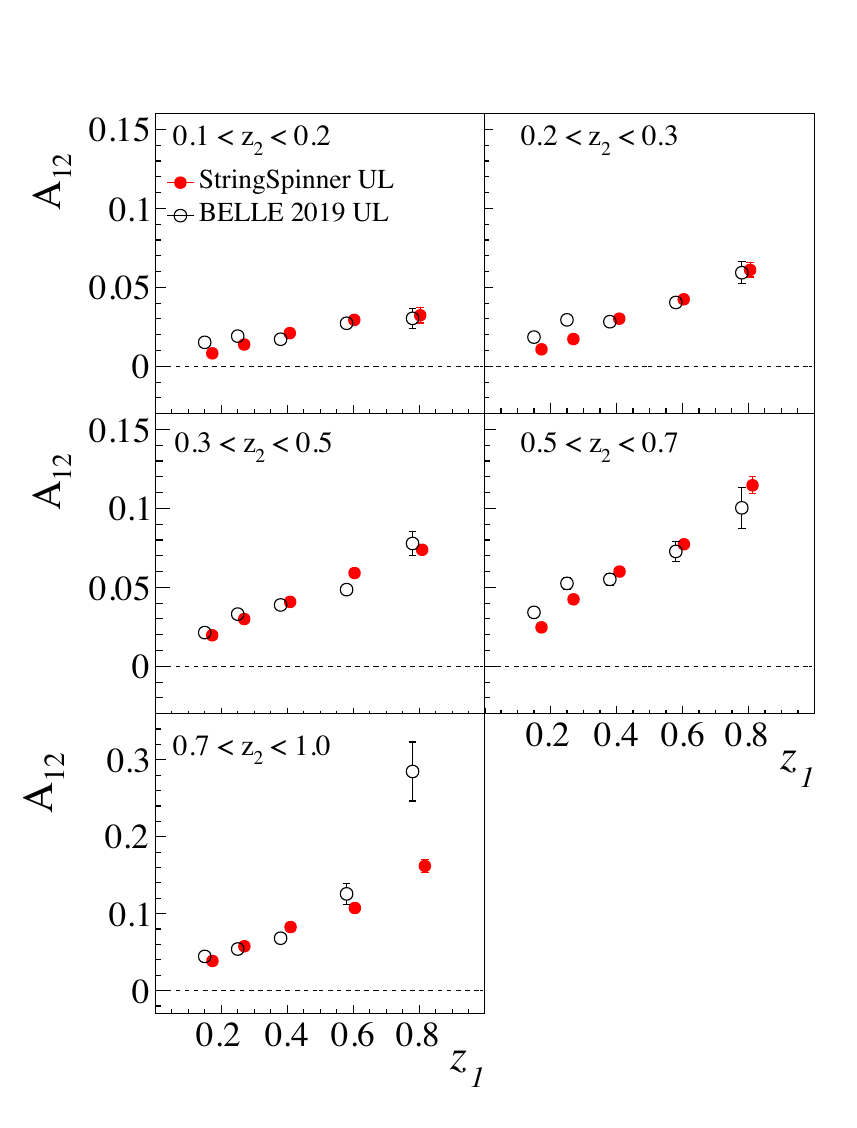}
\end{minipage}
\begin{minipage}[b]{0.42\textwidth}
\includegraphics[width=1.0\textwidth]
{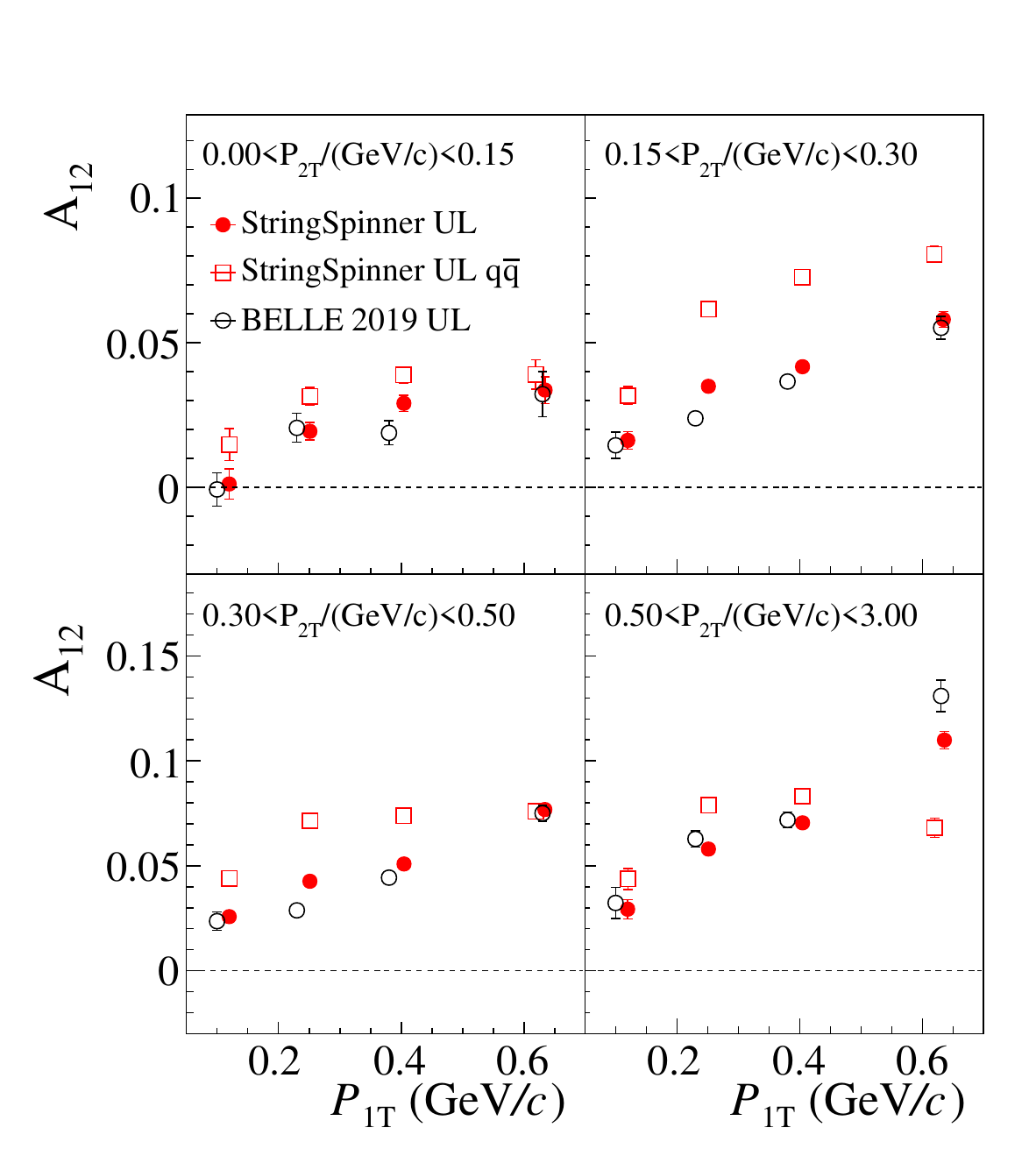}
\end{minipage}
\caption{The Collins asymmetry $\AUL$ for back-to-back $\pi^{\pm}-\pi^{\mp}$ pairs as obtained with StringSpinner (full circles) and as measured by BELLE \cite{Belle:2019nve} (open circles). Top plot: $z_1\times z_2$ binning. Bottom plot: $\PTa\times \PTb$ binning. The open squares represent $\AUL$ evaluated by \StringSpinner{} using the true $\q\qbar$ axis.}
\label{fig:A12 thrust}
\end{figure}

\subsubsection{$\AOneTwo$ asymmetry without thrust axis correction}\label{sec:Results thrust axis}
In this section we study the $\AOneTwoUL$ asymmetry evaluated using the thrust axis $\n$ and not corrected for the smearing effect due to the misalignment between $\n$ and the $\q\qbar$ axis, as done for the BELLE results published in 2019 \cite{Belle:2019nve}. In the simulation, only the fragmentation of $u$, $d$ and $s$ quarks is allowed. The asymmetries are evaluated in the same $z$ and $\PT$ bins used by BELLE~\cite{Belle:2019nve}, and the same cuts $0.1<z_i<1.0$ and $\alpha_i<0.3\,\rm{rad}$ are applied. The latter cut rejects $23\%$ of the hadron pairs. When evaluating the asymmetry as a function of the transverse momentum, each hadron is required to have $z_i>0.2$.

The asymmetry for charged pion pairs as a function of $z_1$ in bins of $z_2$ is shown in the top plot in Fig.~\ref{fig:A12 thrust} (full circles). It exhibits a rising trend as a function of the fractional energy. The size of the asymmetry is smaller as compared to the $\AOneTwoUL$ asymmetry in Fig.~\ref{fig:A12 Belle 2008} evaluated using the true $\q\qbar$ axis. In fact, using the thrust axis the asymmetry is reduced by a factor of about $0.6$ in agreement with the BELLE result \cite{Belle:2008fdv}.

The bottom plot in Fig.~\ref{fig:A12 thrust} shows the $\AOneTwo$ asymmetry as a function of $\PTa$ in bins of $\PTb$. It has nearly a linear trend in the considered transverse momentum range. Considering that the $\AOneTwoUL$ asymmetry is roughly given by the product of two Collins FFs [see Eq.~(\ref{eq:A12})] and the nonmonotonic $\PT$-dependence of the Collins analysing power for the production of pions studied by the standalone simulations in Ref.~\cite{Kerbizi:2021M20}, one would not expect a nearly linear trend for $\AOneTwoUL$ as a function of transverse momentum.
This can be seen by evaluating the asymmetry $\AOneTwoUL$ using the true $\q\qbar$ axis, which is shown by the open squares in the same figure.
Therefore, it turns out that the misalignment between the thrust axis $\n$ and the true $\q\qbar$ axis results in a dilution and a change of the trend of the Collins asymmetry as a function of $\PT$, leading to a nearly linear dependence on this variable.

The corresponding $\AOneTwoUL$ asymmetries measured by BELLE for back-to-back charged pion pairs are shown by the open points in Fig.~\ref{fig:A12 thrust}. Unlike the 2008 BELLE results \cite{Belle:2008fdv}, these asymmetries are not corrected for the contribution of charmed quarks. To obtain the Collins asymmetries in events initiated by $u$, $d$ or $s$ quarks, we multiply the asymmetry measured by BELLE by the factor $(1-f_c)^{-1}$, using the fraction $\fc$ of charm-initiated events given by BELLE and assuming a vanishing Collins asymmetry resulting from such events \cite{Belle:2019nve}.

Looking at the upper plot in Fig.~\ref{fig:A12 thrust}, it can be seen that \StringSpinner{} gives a satisfactory description of the experimental $\AOneTwoUL$ asymmetry as a function of the fractional energy. An exception is the last point for $0.7<z_2<1.0$. From the bottom plot, it can be seen that simulations agree also with the magnitude and the rising trend of the Collins asymmetry as a function of transverse-momentum. 

\begin{figure}[b]
\centering
\begin{minipage}[b]{0.5\textwidth}
\includegraphics[width=1.0\textwidth]{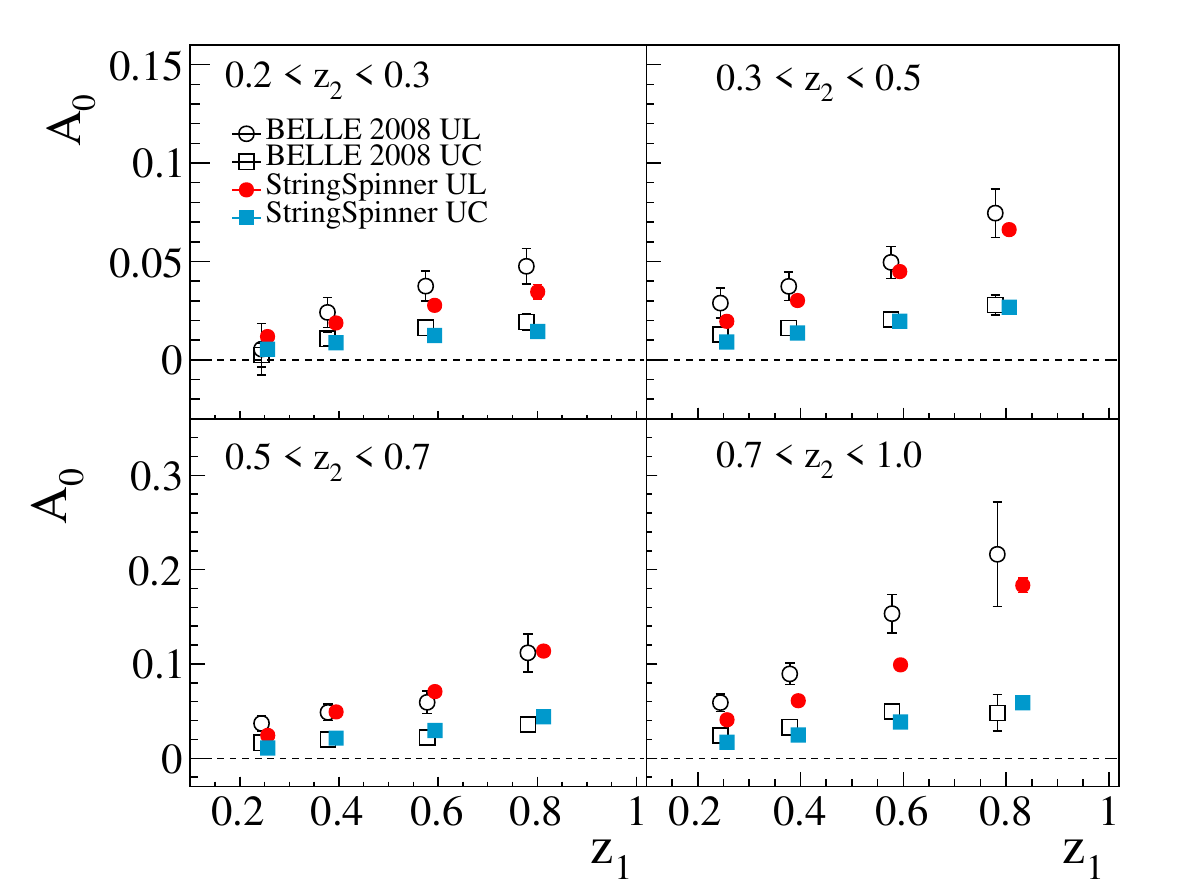}
\end{minipage}
\caption{The Collins asymmetries $\AzeroUL$ (full circles) and $\AzeroUC$ (full squares) for back-to-back $\pi^{\pm}-\pi^{\mp}$ pairs as obtained with \StringSpinner{}, and as measured by BELLE \cite{Belle:2008fdv} (open markers).}
\label{fig:A0 Belle 2008}
\end{figure}

Using \StringSpinner{} with the same parameter settings we evaluated also the Collins asymmetries for back-to-back $\pi^0-\pi^{\pm}$ pairs and $\eta-\pi^{\pm}$ pairs, which were also measured by BELLE in 2019 \cite{Belle:2019nve}. A description of the experimental results similar to that in Fig.~\ref{fig:A12 thrust} was found.

From the point of view of phenomenology, an analysis of the $\AOneTwoUL$ asymmetries measured by BELLE using the thrust axis turns out to be difficult with the presently available theoretical tools.
This would require the cross section for the reaction $e^+e^-\rightarrow h_1\,h_2\,X$ for back-to-back hadrons $h_1$ and $h_2$ including the dependence on the thrust axis, which is not available yet. A step forward in this direction has been performed recently for the reaction $e^+e^-\rightarrow hX$, the cross section of which has been applied to the study of the transverse momentum dependence of the spin-averaged $D_{1q}^h$ FF \cite{Boglione:2023duo}.

\begin{figure}[tbh]
\centering
\begin{minipage}[b]{0.5\textwidth}
\includegraphics[width=1.0\textwidth]{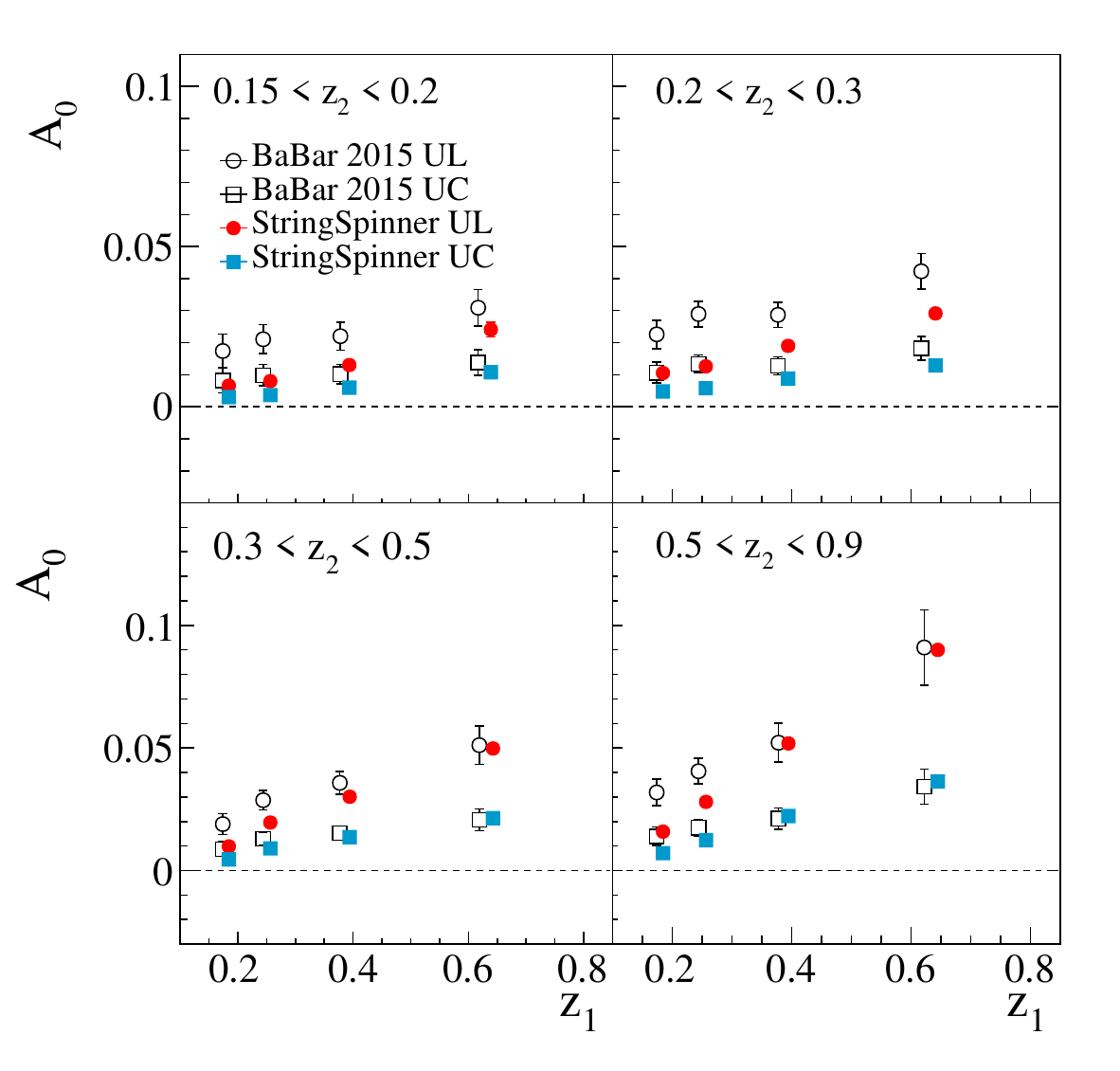}
\end{minipage}
\caption{The Collins asymmetries $\AzeroUL$ (full circles) and $\AzeroUC$ (full squares) for back-to-back $\pi^{\pm}-\pi^{\mp}$ pairs as obtained with \StringSpinner{}, and as measured by BABAR \cite{BaBar:2015mcn} (open markers).}
\label{fig:A0 BaBar 2014}
\end{figure}

\subsection{Results on the $\Azero$ asymmetry}\label{sec:Hadronic Method}
To calculate the Collins asymmetry $\Azero$ we have used the same simulated events used to obtain the $\AOneTwo$ asymmetry of Sec.~\ref{sec:Thrust Method}. The $\Azero$ asymmetry has also been corrected by the experiments to account for the use of the thrust axis instead of the true $\q\qbar$ axis. This asymmetry is however less sensitive to the misalignment between $\n$ and the $\q\qbar$ axis, the correction factor being about $1.10$ \cite{Belle:2008fdv,BaBar:2013jdt}.

\paragraph{Comparison with the BELLE data.} The results for the asymmetries $\AzeroUL$ (full circles) and $\AzeroUC$ (full rectangles) for back-to-back charged pions are shown in Fig.~\ref{fig:A0 Belle 2008} as a function of $z_1$ for selected bins of $z_2$. The kinematic selections and the $z$-binning are the same as in the BELLE 2008 analysis in Ref.~\cite{Belle:2008fdv} (and as in Fig.~\ref{fig:A12 Belle 2008}). As can be seen, the $\Azero$ asymmetry also exhibits a rising trend as a function of $z$. Comparing to the $\AOneTwo$ asymmetry in Fig.~\ref{fig:A12 Belle 2008}, $\Azero$ turns out to have lower values. This is expected because the use of the momentum of $h_2$ to build the hadron frame in Fig.~\ref{fig:kinematic M0} introduces a smearing analogous to the thrust axis. By comparing Fig.~\ref{fig:A0 Belle 2008} with Fig.~\ref{fig:A12 thrust}, it is interesting to note that the size of the $\Azero$ asymmetry is similar to that of the $\AOneTwo$ asymmetry without the correction for the thrust axis smearing.

The $\Azero$ asymmetries measured by BELLE \cite{Belle:2008fdv} are also shown in Fig.~\ref{fig:A0 Belle 2008} (open markers). As can be seen, \StringSpinner{} gives a satisfactory description of the trend as a function of the fractional energy and of the magnitude of both the $\AzeroUL$ and $\AzeroUC$ asymmetries. An exception is the interval $0.7<z_2<1.0$, where the simulated results for the $\AzeroUL$ asymmetry is systematically lower than the BELLE data. This is not the case for the $\AzeroUC$ asymmetry.

\begin{figure}[b]
\centering
\begin{minipage}[b]{0.5\textwidth}
\includegraphics[width=1.0\textwidth]{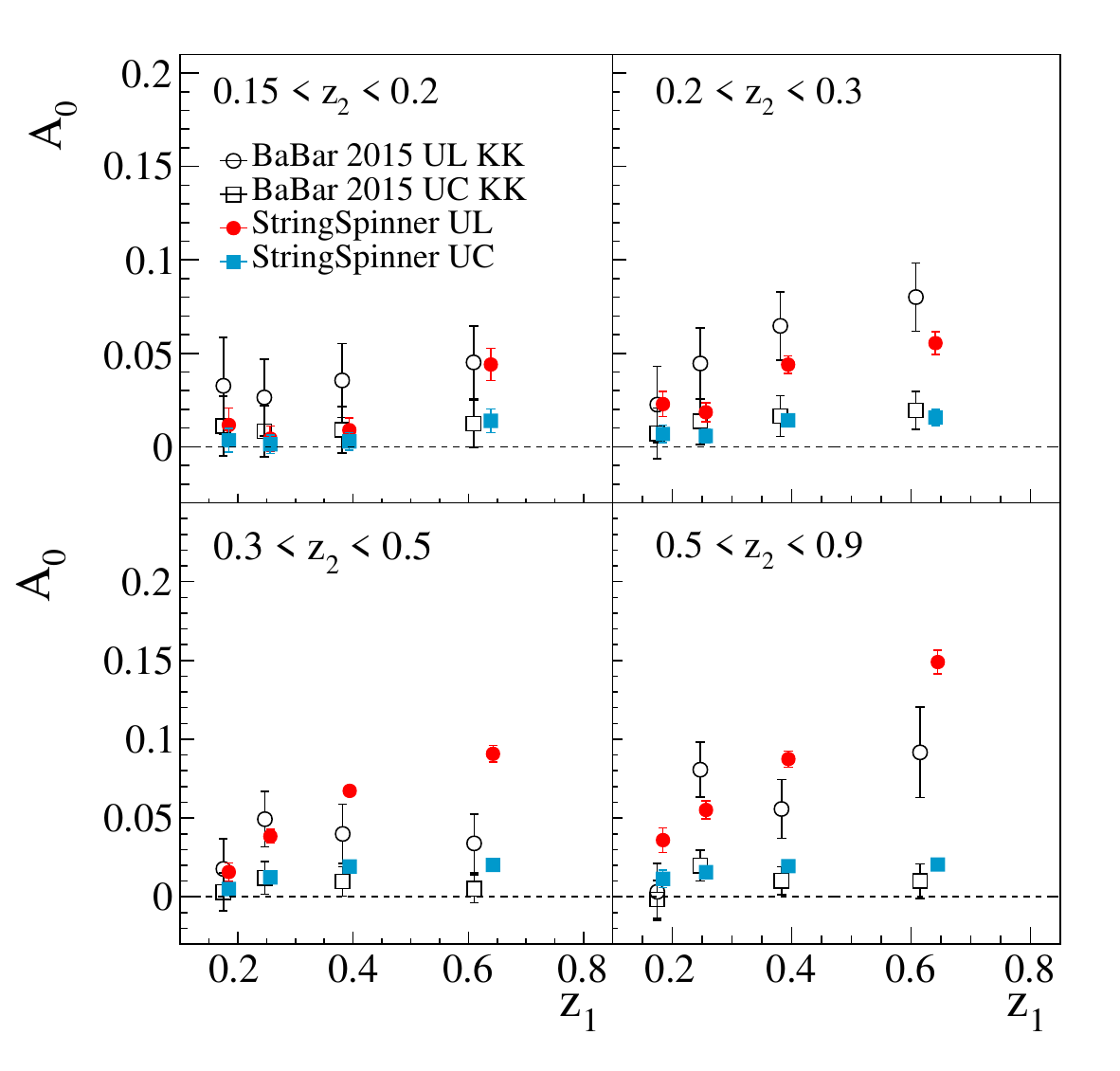}
\end{minipage}
\caption{The Collins asymmetries $\AzeroUL$ (full circles) and $\AzeroUC$ (full squares) for back-to-back $K^{\pm}-K^{\mp}$ pairs as obtained with \StringSpinner{}, and as measured by BABAR \cite{BaBar:2015mcn} (open markers).}
\label{fig:A0 KK BaBar 2015}
\end{figure}

\paragraph{Comparison with BABAR data.} In Fig.~\ref{fig:A0 BaBar 2014}, we show the simulated $\Azero$ asymmetry (closed points) in the same two-dimensional $z_1\times z_2$ binning as in the 2015 BABAR analysis \cite{BaBar:2015mcn}. The trend and the size of both $\AzeroUL$ and $\AzeroUC$ are similar to what obtained for BELLE in Fig.~\ref{fig:A0 Belle 2008}. This is expected since the annihilation events are carried at the same CMS energy, and that the selections applied in the analysis are similar (the cut $\alpha_i<\pi/4$ applied by BABAR rejects only the $9\,\%$ of the simulated data). The corresponding Collins asymmetries measured by BABAR in 2015 are shown by the open markers in Fig.~\ref{fig:A0 BaBar 2014}. A satisfactory description of the $\AzeroUL$ asymmetry is found for $z_1>0.3$, while if at least one of the fractional energies is less than $0.3$ the simulated asymmetries are lower than the data.
A similar description is also found for the $\AzeroUC$ asymmetry.

In Fig.~\ref{fig:A0 KK BaBar 2015} are shown the simulation results for the $\AzeroUL$ (full circles) and $\AzeroUC$ (full squares) asymmetries for charged kaons, in the same binning for the fractional energy as in the BABAR analysis \cite{BaBar:2015mcn}. The asymmetry has a rising trend as a function of the fractional energy and it is lower than the $\AOneTwo$ asymmetry for charged kaons shown in Fig.~\ref{fig:A12 KK BaBar 2015}. This is analogous to the $\Azero$ asymmetry for charged pions.

The corresponding asymmetries measured by BABAR are shown in Fig.~\ref{fig:A0 KK BaBar 2015} by the open markers. Despite the large uncertainties in the data, the agreement between \StringSpinner{} and the data is satisfactory for both the $\AzeroUL$ and $\AzeroUC$ asymmetries.

\begin{figure}[h]
\centering
\begin{minipage}[b]{0.5\textwidth}
\includegraphics[width=0.85\textwidth]{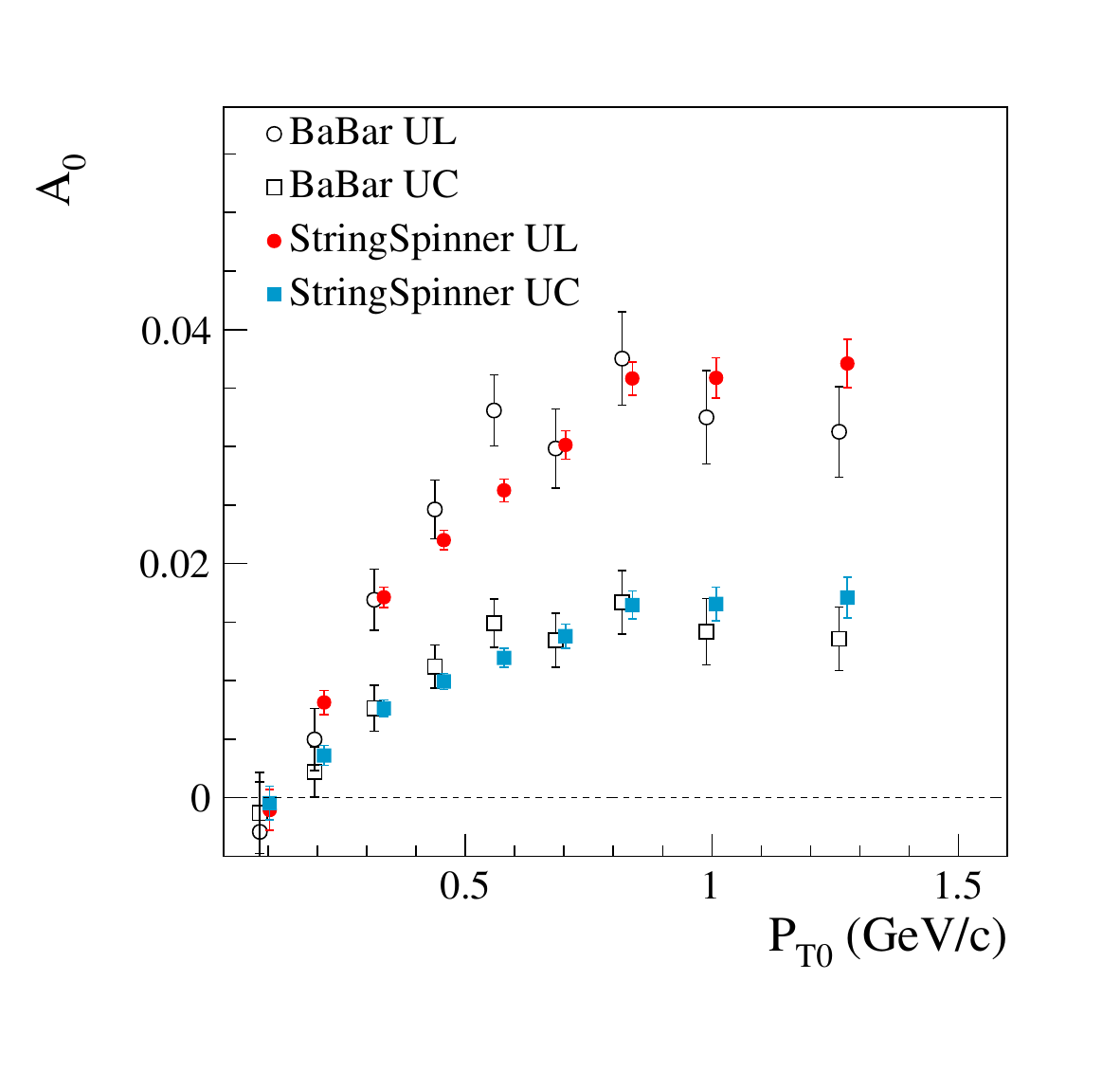}
\end{minipage}
\caption{Comparison between the Collins asymmetries $\AzeroUL$ (full circles) and $\AzeroUC$ (full squares) as a function of $\PTzero$ for back-to-back $\pi^{\pm}-\pi^{\mp}$ pairs obtained with \StringSpinner{} (full markers), and the asymmetries measured by BABAR \cite{BaBar:2013jdt} (open markers).}
\label{fig:A0 PT0 BaBar 2014}
\end{figure}

Finally, to further investigate the transverse-momentum dependence of the $\Azero$ asymmetry predicted by the string+${}^3P_0$ model it is interesting to evaluate the $\AzeroUL$ and $\AzeroUC$ asymmetries for charged pion pairs as a function of $\PTzero$ using the simulated events. The simulation results for $\AzeroUL$ (closed circles) and $\AzeroUC$ (closed squares) are shown in Fig.~\ref{fig:A0 PT0 BaBar 2014}. Both asymmetries have a peculiar transverse momentum dependence. They show a rising trend as a function of $\PTzero$ for $\PTzero\leq 0.7\,\GeV/c$ while for larger transverse momenta they flatten out. This is seen also in the corresponding BABAR results \cite{BaBar:2015mcn} shown in the same figure by the open markers. As can be seen, the agreement between the simulations and the experimental data is remarkable both for the trend and the size of the asymmetries.

This result and the satisfactory description of the transverse-momentum dependence of $\AOneTwoUL$ evaluated using the thrust axis in Fig.~\ref{fig:A12 thrust} (bottom), represent a test of the transverse-momentum-dependent effects in $e^+e^-$ annihilation predicted by the string+${}^3P_0$ model. The reproduction of these effects is encouraging and motives further developments of the model.

\section{The Collins analysing power in the string+${}^3P_0$ model}\label{sec:analysing-power}
As can be seen from Eqs.~(\ref{eq:A12}) and (\ref{eq:A0}) the essential ingredient of the asymmetries $\AOneTwo$ and $\Azero$ is the ratio between the Collins function $\Hq$ and the spin-averaged FF $D_{1q}^h$. This ratio is related to the Collins analysing power for the fragmentation of $q$ in $h$,
\begin{equation}\label{eq:AP}
\Ap(z,\ptabs)=\frac{\ptabs}{z\,m_h}\,\frac{\Hq(z,\ptabs)}{\Dq(z,\ptabs)},
\end{equation}
which gives information on the nonperturbative spin-dependent dynamics involved in the hadronization process. It depends on $z$ and the transverse momentum $\ptabs$ of $h$ with mass $m_h$. 

The Collins analysing power is also obtained from phenomenological analyses aimed at extracting the transversity PDF and the Collins FF from the combined analysis of SIDIS and $e^+e^-$ data \cite{Anselmino:2015fty,Martin:2014wua,Anselmino:2015sxa,Kang_transv_evolution,Cammarota:2020qcw}. In such analyses a parametrization is chosen for the $z$-dependence of $\Hq$, while the Gaussian ansatz is assumed for the $\ptabs$-dependence. These assumptions are then reflected on the dependence of the Collins analysing power on $z$ and $\ptabs$. Furthermore, assuming isospin and charge conjugation invariance, the favoured spin-averaged FF
$\DF$=$D_{1u}^{\pi^+}$=$D_{1d}^{\pi^-}$=$D_{1\bar{u}}^{\pi^-}$=$D_{1\bar{d}}^{\pi^+}$ and the unfavoured spin-averaged FF $\DU$=$D_{1u}^{\pi^-}$=$D_{1d}^{\pi^+}$=$D_{1\bar{u}}^{\pi^+}$=$D_{1\bar{d}}^{\pi^-}$ are introduced. Analogously, the favoured ($\HF$) and unfavoured ($\HU$) Collins FFs are defined. Using these definitions, one obtains the favoured (unfavoured) Collins analysing power $\ApF$ ($\ApU$) by inserting $\HF$ and $\DF$ ($\HU$ and $\DU$) in Eq.~(\ref{eq:AP}).

The string+${}^3P_0$ model, on the other hand, produces its own prediction for the Collins analysing power. This quantity was extensively studied in the previous works by the means of a standalone Monte Carlo implementation of the string+${}^3P_0$ model \cite{Kerbizi:2018qpp,Kerbizi:2019ubp,Kerbizi:2021M20}. However, given the satisfactory description of the $e^+e^-$ data achieved in Sec.~\ref{sec:Results}, it is interesting to evaluate the resulting Collins analysing power and compare it with phenomenological analyses. After recalling the steps for calculating the Collins analysing power from simulated data in Sec.~\ref{sec:analysing-power-calculation}, we show the results and the comparison with phenomenological analyses in Sec.~\ref{sec:Comparison analysing power}.

\subsection{Calculation of the Collins analysing power from simulated data}\label{sec:analysing-power-calculation}
The Collins analysing power $\Ap$ can be accessed in simulations of the fragmentation chain of a transversely polarized quark $q^{\uparrow}$. In the string+${}^3P_0$ model it consists in simulating the fragmentation of a string stretched between a $\q^{\uparrow}\qbar$ pair where only $q$ is transversely polarized and the fragmentation chain is evolved from $q$ towards the $\qbar$ side. Taking the transverse polarization of $q$ with respect to the string axis to be $\SqT$, the joint spin density matrix to be used in simulations is given by $\rho(\q,\qbar)=\rho(q)\otimes \Iqbar$, where the spin density matrix of $q$ is $\rho(q)=(\Iq+\sigmaq\cdot\SqT)/2$. To simplify, we take $q$ to be in the pure state with $\SqT=\yq$ ($|\SqT|=1$). 

The azimuthal distribution of the hadron $h$ produced in the fragmentation of $q^{\uparrow}$ is expected to be \cite{Collins:1992kk}
\begin{eqnarray}\label{eq:Nh}
    \frac{dN_h}{d\phi_C\,d\,z\,d\ptabs}\propto1+\Ap(z,\ptabs)\,|\SqT|\,\sin\phiC,
\end{eqnarray}
where the Collins angle $\phiC$ is given by $\phiC=\phiS-\phi_h$. $\phiS$ and $\phi_h$ are the azimuthal angle of $\SqT$ and the azimuthal angle of $\pt$, respectively, evaluated in the QHF. The Collins analysing power is thus the amplitude of the $\sin\phiC$ modulation. As can be deduced from Eq.~(\ref{eq:Nh}), it can be calculated as $\Ap(z,\ptabs)=2\langle \sin\phiC\rangle$ in a selected $z$ or $\ptabs$ interval.

To calculate the Collins analysing power we performed simulations of fragmentations of $u^{\uparrow}\bar{u}$ strings at the CMS energy $\sqrt{s}=10.6\,\GeV$ using the same free parameters as in Sec.~\ref{sec:setting}.

\subsection{Comparison of the Collins analysing power with phenomenological analyses}\label{sec:Comparison analysing power}
As a preliminary step, we checked that the isospin and charge conjugation relations used to define the favoured and unfavoured FFs hold also in simulations. This is as expected because the isospin and charge conjugation symmetries are employed in \Pythia{} to calculate the probabilities for projecting a given quark-antiquark pair onto a hadronic state. 

\begin{figure}[tbh]
\centering
\begin{minipage}[b]{0.40\textwidth}
\includegraphics[width=0.99\textwidth]{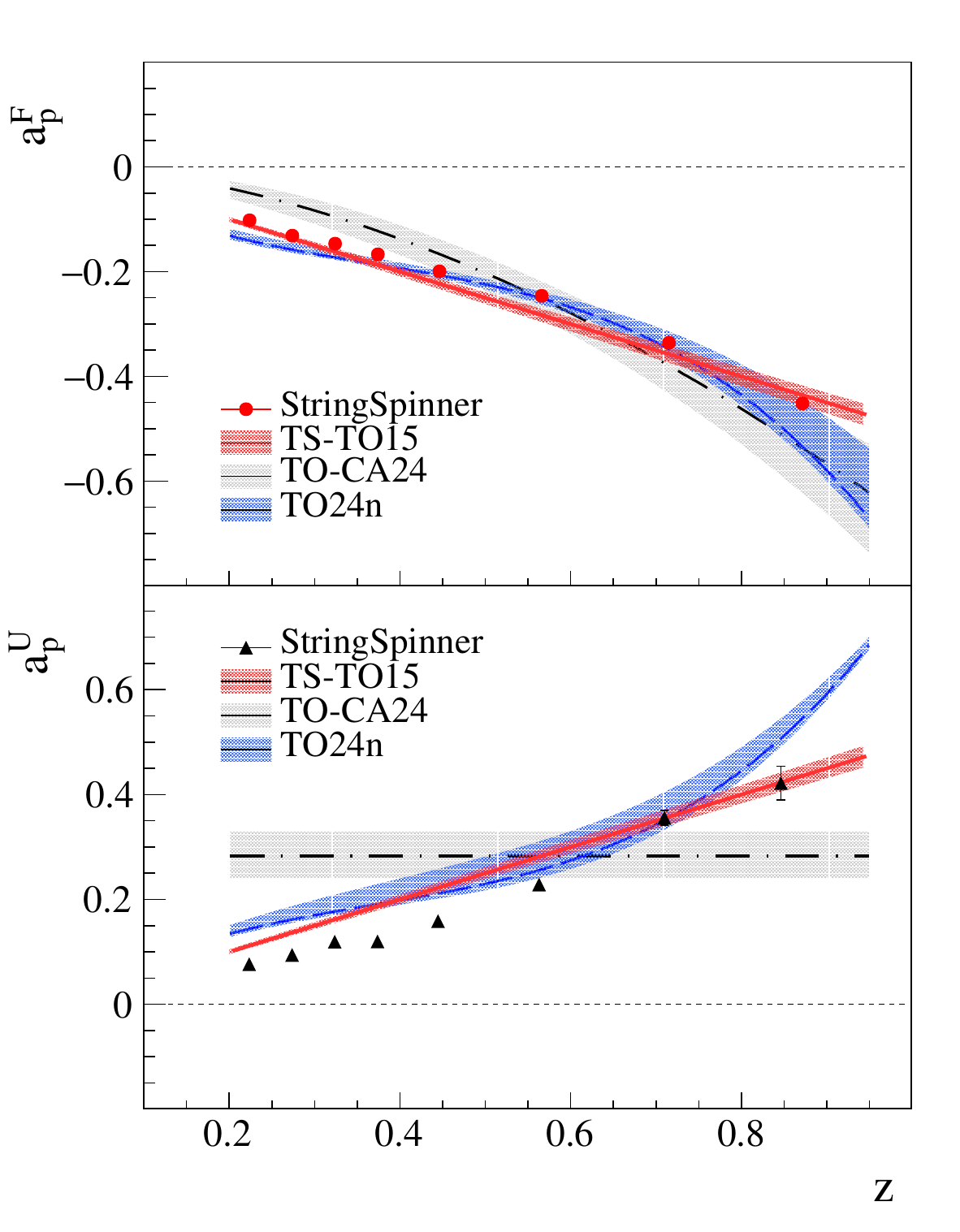}
\end{minipage}
\caption{Comparison between the Collins analysing power for favoured (circles) and unfavoured (triangles) fragmentation as obtained from the string+${}^3P_0$ model and the corresponding results from different phenomenological analyses (see text).}
\label{fig:Collins ap}
\end{figure}

Afterwards we evaluate $\ApF$ and $\ApU$ by looking at the distribution in Eq.~(\ref{eq:Nh}) for, respectively, final state $\pi^+$ and $\pi^-$ mesons. The result is shown as a function of $z$ in Fig.~\ref{fig:Collins ap}. The top panel shows $\ApF$ as given by the string+${}^3P_0$ model (circles), while the bottom panel shows $\ApU$ (triangles). The favoured and unfavoured analysing powers have about the same size but opposite sign~\cite{Kerbizi:2018qpp,Kerbizi:2019ubp,Kerbizi:2021M20}.

In the same figure are shown the favoured and unfavoured Collins analysing powers evaluated by using the Collins FFs extracted by different groups \cite{Martin:2014wua,Boglione:2024dal,Boglione:2024new}. The same color-code is used for both $\ApF$ and $\ApU$. 
The continuous line is the analysing power extracted in Ref.~\cite{Martin:2014wua} (TS-TO15) using only the BELLE data on $\AOneTwoUL$ \cite{Belle:2008fdv} and assuming the relation $\ApF=-\ApU$ ("scenario 2"). The enveloping red band represents a two standard deviations confidence interval (CL) evaluated by using the estimated uncertainty on the only free parameter of the $\ApF$ \cite{Martin:2014wua}. The gray band shows the 95\% CL interval for the analysing power from the analysis performed in Ref.~\cite{Boglione:2024dal} (TO-CA24) using the SIDIS, $e^+e^-$ and proton-proton data. The dashed-dotted line is the corresponding median value of the analysing power. The blue band shows the analysing power as obtained by a revisited TO-CA24 analysis (TO24n) \cite{Boglione:2024new} that in addition includes the recent Collins asymmetries in SIDIS with a deuteron target measured by COMPASS \cite{COMPASS:2023vhr} and assumes a polynomial dependence on the Collins analysing power on $z$ as in Ref.~\cite{Anselmino:2013vqa}. The band represents a two standard deviations CL interval, while the dashed line is the mean value of the analysing power.

As can  be seen from the top panel in Fig.~\ref{fig:Collins ap} the favoured Collins analysing power increases as a function of $z$, a feature common to the different extractions. The string+${}^3P_0$ result is remarkably similar to the TO24n extraction and not too far from a linear dependence on $z$ as in the TS-TO15 extraction. Compared to the TO-CA24 extraction, the $\ApF$ results are similar for $z>0.3$ but differ for smaller values. Concerning the unfavoured analysing power $\ApU$, the string+${}^3P_0$ model predicts again a rising trend with $z$, as also obtained in TS-TO15 and TO-CA24. 

A different result for $\ApU$ is obtained in the TO-CA24 extraction, where a constant expression for the analysing power as a function of $z$ (i.e., same $z$-dependence for $\HU$ and $\DU$) is found to be adequate for a satisfactory description of the data. This assumption has been successfully used also in previous analyses \cite{Anselmino:2015sxa,DAlesio:2020vtw}. 
The difference between the two results can be seen as an additional uncertainty on the knowledge of the unfavoured Collins FF as obtained from phenomenological analyses.

In this respect, models of polarized hadronization, such as the string+${}^3P_0$ model, can be used as a guide for the choice of the parametrization of the Collins FFs to be employed in the phenomenological analysis of SIDIS, $e^+e^-$ and proton-proton data. In particular, the string+${}^3P_0$ model suggests a polynomial functional form for the $z$ dependence of the favoured and unfavoured Collins analysing powers. It implements the fact that the information on the spin state of the fragmenting quark decays along the fragmentation chain and the memory of this state becomes negligible at small $z$.

\section{Conclusions}\label{sec:Conclusions}
The string+${}^3P_0$ model of hadronization is implemented for the first time in the \Pythia{} \texttt{8} event generator for the simulation of the $e^+e^-$ annihilation to hadrons with quark spin effects. We use the recursive recipe for the simulation of the spin-dependent string fragmentation of a quark-antiquark pair with entangled spin states proposed recently in Ref.~\cite{Kerbizi:2023luv}. The actual implementation in \Pythia{} is performed by developing further the \StringSpinner{} package, which now can be applied to generate either polarized DIS events or $e^+e^-$ events.

In this work, we used the new package to simulate $e^+e^-$ annihilation events at the CMS energy $\sqrt{s}=10.6\,\GeV$ assuming the annihilation to occur by the exchange of a virtual photon. This corresponds to the kinematic configuration of the BELLE and BABAR experiments. The generated events are analyzed to calculate the Collins asymmetries for back-to-back hadrons in the $e^+e^-$ CMS system by exploiting both the thrust axis method and the hadronic plane method. The simulated Collins asymmetries are compared to data from the BELLE and BABAR collaborations. A satisfactory comparison is found with the BELLE data for the both Collins asymmetries calculated by using the thrust axis method and the hadronic plane method. The comparison with the BABAR data is also satisfactory, with the exception of the $\AOneTwo$ asymmetries for charged pions. The inconsistencies between the BABAR and the BELLE results for these asymmetries are, however, known.

Using the simulations, we also calculated the favoured and unfavoured Collins analysing powers predicted by the string+${}^3P_0$ model as a function of the fractional energy. The results were shown to be similar to those of phenomenological extractions that assume a non-constant dependence of the unfavoured analysing power on the fractional energy.

To conclude, considering the encouraging results obtained in this work, we believe that the string+${}^3P_0$ model is a sound model for the description of the quark-spin effects in hadronization and that it can be used for a systematical implementation of such effects in Monte Carlo event generators.

The \StringSpinner\ code used in this paper will become available as a contributed module to \Pythia, available from gitlab at  \texttt{gitlab.com/pythia8-contrib}.

\begin{acknowledgments}
The authors are grateful to Xavier Artru for the many enlightening discussions, to Isabella Garzia for clarifications on the BABAR data, and to Elena Boglione and Carlo Flore for useful discussions on the parametrizations of the Collins function and for providing the bands of the TO-CA24 and TO24n fits.

The work of A.~K.~is done in the context of the project “POLFRAG:
Simulation of polarized quark fragmentation and application to the
investigation of the nucleon structure", CUP No. J97G22000510001,
funded by the Italian Ministry of University and Research
(MUR). A.~K.~acknowledges also support from the University of Trieste via
the project “Simulazione delle correlazioni quantistiche in collisioni
ad alta energia", CUP No. J93C22001380002. L.~L.~was supported by the
MCnetITN3 H2020 Marie Curie Initial Training Network, contract
722104. In addition L.~L.~is supported by Swedish Research Council
contract 2020-04869.
\end{acknowledgments}



\bibliography{apssamp}

\end{document}